\documentclass[aps,prl,twocolumn,superscriptaddress]{revtex4-1}

\usepackage[pdftex]{graphicx}
\usepackage{amsmath}
\usepackage[amssymb,Gray]{SIunits}
\usepackage{xfrac}
\usepackage{tikz}

\usepackage{graphicx}
\usepackage{dcolumn}
\usepackage{bm}

\begin{document}


\title{Parametric Feedback Cooling of Levitated Optomechanics in a Parabolic Mirror Trap}

\author{Jamie Vovrosh}
\affiliation{Department of Physics and Astronomy, University of Southampton, SO17 1BJ, United Kingdom}
\author{Muddassar Rashid}
\affiliation{Department of Physics and Astronomy, University of Southampton, SO17 1BJ, United Kingdom}
\author{David Hempston}
\affiliation{Department of Physics and Astronomy, University of Southampton, SO17 1BJ, United Kingdom}
\author{James Bateman}
\affiliation{Department of Physics and Astronomy, University of Southampton, SO17 1BJ, United Kingdom}
\affiliation{Department of Physics, College of Science, Swansea University,
Swansea SA2 8PP, UK}
\author{Mauro Paternostro}
\affiliation{Centre for Theoretical Atomic, Molecular, and Optical Physics,
School of Mathematics and Physics, Queen's University, Belfast BT7 1NN, United Kingdom}
\author{Hendrik Ulbricht}
\email[]{H.Ulbricht@soton.ac.uk}
\affiliation{Department of Physics and Astronomy, University of Southampton, SO17 1BJ, United Kingdom}


\begin{abstract}
Abstract:  Levitated optomechanics, a new experimental physics platform, holds promise for fundamental science and quantum technological sensing applications. We demonstrate a simple and robust geometry for optical trapping in vacuum of a single nanoparticle based on a parabolic mirror and the optical gradient force. We demonstrate rapid parametric feedback cooling of all three motional degrees of freedom from room temperature to a few mK. A single laser at 1550nm, and a single photodiode, are used for trapping, position detection, and cooling for all three dimensions. Particles with diameters from 26nm to 160nm are trapped without feedback to 10$^{-5}$mbar and with feedback-engaged the pressure is reduced to 10$^{-6}$mbar. Modifications to the harmonic motion in the presence of noise and feedback are studied, and an experimental mechanical quality factor in excess of 4$\times 10^7$ is evaluated. This particle manipulation is key to build a nanoparticle matter-wave interferometer in order to test the quantum superposition principle in the macroscopic domain.\\
\\


\end{abstract}

\maketitle


{\bf Introduction:} Levitated optomechanics, where light is used to trap a single particle and manipulate its centre-of-mass motion, holds promise for unprecedented precision in the control of massive nanoparticles in vacuum~\cite{RevModPhysCavityoptomechanics}. Such precision will be key for the development of applications in sensing of magnetic spin resonance~\citep{Levispin}, investigating gravitational effects and detecting inertial forces.  Background-free spectroscopy experiments with single particles will benefit from such exquisite control. Moreover, new avenues through which to test fundamental science will become possible, such as interferometric and non-interferometric tests of the quantum superposition principle~\cite{macro} as well as the test of the interplay between quantum theory and gravity ~\cite{RevModPhysBassi, Gravitation, Penrose1996581,ISI000358741800023}. Recently the motion of single micro/nanoparticles has been optically detected and cooled by linear~\cite{Li2011} and parametric feedback~\cite{PhysRevLett.109.103603}. One experiment has already reached the single photon recoil limit~\cite{Jain2016}. The motion of neutral particles has been cooled by cavity-assisted methods~\cite{Asenbaum2013, kiesel2013cavity, Chang19012010}. For charged particles, the use of Paul traps combined with optical cavities has led to temperatures below one Kelvin in high vacuum ($\simeq10^{-6}$mbar)~\cite{millen2015cavity}. Elsewhere, the rotation and torsional motion of nanostructures has been studied and controlled~\cite{arita2013laser, kuhn2015cavity, PhysRevLett.117.123604}. Further the first demonstration of squeezing the classical thermal state has been reported~\cite{rashid2016experimental}.

\begin{figure*}
    \centering
    \includegraphics[width=0.95\textwidth]{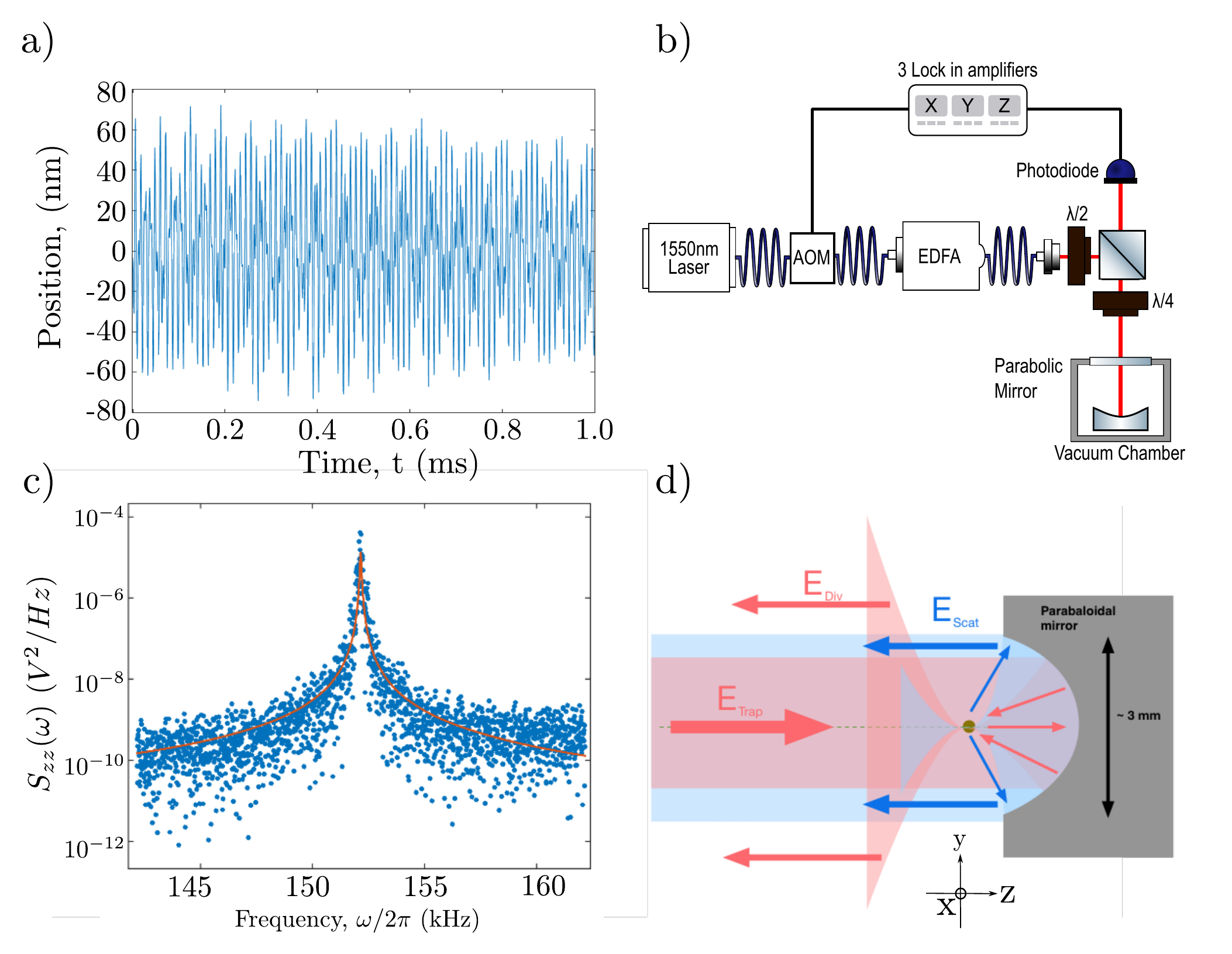}
    \caption{\textbf{Trapping and controlling the motion of the particle.} 
    a) Time trace of the particle's three-dimensional motion in the trap, the total signal measured by a single photodiode detector. The particle's motion along the ($x$,$y$,$z$)-directions modulates the phase of the backscattered light re-collimated by the parabolic mirror. The modulated light is measured by a single photodiode and the position is calculated from the fit in Fig.1c) (see supplement S1 for details). No feedback is applied and the pressure is 7$\times$10$^{-2}$mbar. While the spring constant for $x$ and $y$ are symmetric, the respective spectral peaks are separated by manipulation of the polarisation of the trapping laser. 
    b) Optical setup for parametric feedback cooling. The three-dimensional position of a trapped nanoparticle is measured using a single photodiode. The signal is {\it frequency doubled} and {\it phase shifted} using three lock-in amplifiers.  The sum of these signals is used to modulate the intensity of the trapping laser by an {\it acusto optical modulator} (AOM). The modulated signal is amplified in an optical {\it Erbium doped fibre amplifier} (EDFA) (see supplement S6 for more details). 
    c) Power spectral density (PSD) of time trace shown in a). The plot shows the zoom into the motion in $z$-direction. The PSD is fitted to Eq.(\ref{eq:PSD}), which is used to extract information about the trapped particle and its motion (see supplement S1 for details).
    d)  We show the incident and outgoing light fields at the parabolic mirror. The detection of the position of the particle is using optical interferometry between the light scattered off the particle $E_{\rm scat}$ and the diverging reference $E_{\rm div}$. Here the position of the detector is chosen so that the amplitudes of both fields are comparable, unlike the standard homodyne detection scheme (c.f. supplement S2 for details on position resolution of this interferometric scheme).}
    \label{fig1}
\end{figure*}

In this paper, we demonstrate a single-mirror trap geometry which has sufficient position resolution to approach a low-phonon state of the centre-of-mass motion of the trapped nanoparticle by parametric feedback cooling alone. Through this mechanism, we have been able to reduce the amplitude of the motion of the centre-of-mass of the particle to energies corresponding to effective temperatures of just 1mK. Correspondingly, our trapped nanoparticle is prepared in motional states thermally populated by $\sim150$ phonons in average. 

{\bf Results:} The particle is trapped by the optical gradient force, and detected by light scattered in the Rayleigh regime (cf. Fig.~\ref{fig1})) for an illustration of the experimental setting).
The reflective geometry used in our experiment allows for very tight focusing via a large numerical aperture, which minimises the optical scattering force (radiation pressure) for sub-micron particles. In the limit where the particle is undergoing small-amplitude oscillations, the trapping potential is harmonic and the motion along the three spatial directions (labelled as $x$, $y$, and $z$) is decoupled. Under these conditions, motion along a given spatial direction can be characterised by a trap frequency $\omega_{0,\{z,x,y\}}= \sqrt{k_{0,\{z,x,y\}}/m} $, which is defined by the particle mass $m$ and the optical spring constant $k_{0,z}={2\alpha P}/({c \pi \epsilon_0 \rm w_{f}^{6} /\lambda^2})$ ($z$-direction) and a different spring constant in the $x,y$ directions due to the different gradient of the Gaussian light focus: $k_{0,\{x,y\}}={8 \alpha P}/({c \pi \epsilon_0 \rm w_{f}^{4}})$ for a polarisable particle that is dipole-trapped in a focused Gaussian laser beam. Here $P$ is the trapping laser power, $\alpha$ the polarisability of the particle, $\rm w_f$ the waist of the trapping laser beam at the focus, $c$ the speed of light, $\lambda$ is the wavelength of the trapping laser and $\epsilon_0$ the permittivity of vacuum. For large amplitude oscillations, behaviour is anharmonic and motional modes are coupled, as has been observed in both low vacuum and high vacuum~\cite{Gieseler2013}.

We control the particle motion by feedback modulation of the optical spring constant, which is the control parameter of the motion, via variation of the trapping laser power. We define the depth of this modulation as $\eta = I_{fb}/I_{0}$where $I_{0}$ is the laser intensity without feedback and $I_{fb}$ is the amplitude of the feedback modulation. Such intensity modulation provides a time-varying, non-conservative motional damping resulting in {\it small} oscillation amplitudes. The equation of motion for the trapped particle in the $x$-direction is
\begin{equation}\label{eq:particlemotion}
\ddot{x}(t)+\Gamma_{0} \dot{x}(t) +\frac{k_{0} +k_{fb}(t)}{m}x(t)= \frac{F_{\rm th}(t)}{m} ,
\end{equation}
where $\Gamma_0$ is the damping rate of the particle's motion. We have introduced $F_{\rm th}$ as a stochastic driving term arising from those same sources origin to the motional damping $\Gamma_0$ - at the parameter range of this experiment dominated by collisional mechanisms - and $k_{fb}(t)$ as the variation in trap stiffness caused by parametric feedback. Similar equations hold for the particles motion in $y$ and $z$ directions.
\begin{figure*}
    \centering
    \includegraphics[width=0.95\textwidth]{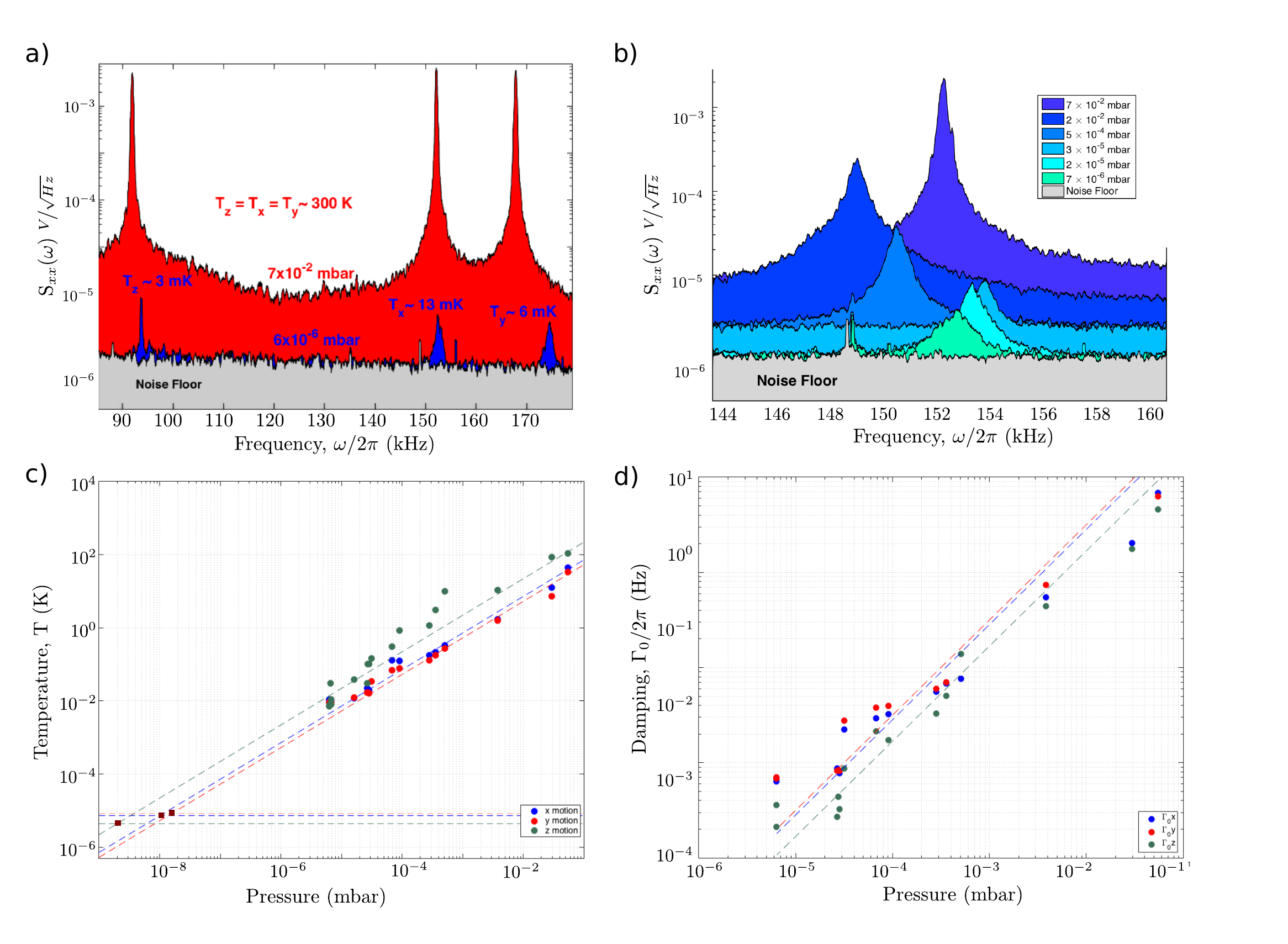}
    \caption{\textbf{Analysing cooling rate $\delta\Gamma$ and damping rate $\Gamma_0$.} 
  a) We show an instance of PSD with $x$, $y$, $z$ peaks at temperatures $T_{cm}$ taken simultaneously. The red data represents a PSD without feedback at 300 K at a pressure of $7\times10^{-2}$ mbar. The blue data is a PSD of a feedback cooled particle at $6\times10^{-6}$ mbar. The lowest temperatures reached in this experiment are indicated in the blue graph. The motion in $x$, $y$, $z$-directions are cooled with independent feedback parameters. The grey data represents the noise floor of our system. 
  b) Shows a more detailed variation of the feedback cooling of $y$ at different pressures. Feedback parameters are optimised to maximise the cooling effect. We observe a frequency shift for PSDs taken at different motional temperature $T_{cm}$ and background pressure, $p$. 
  c) Shows the translational temperature $T_{cm}$ for the motion in $x$, $y$, $z$-direction vs pressure, $p$. An extrapolation of the fit (shown in dashed-lines) through the cooling rate shows that for this data set the ground state temperature (intercept with horizontal dashed lines) could be reached for the pressures of 10$^{-8}$ to 10$^{-9}$ mbar; if photon recoil and other sources of noise can be overcome~\cite{Jain2016}. 
  d) The damping rate $\Gamma_0$ as extracted from fitting a Lorentzian to the PSD (see supplement S1 for details), here is shown for $x$, $y$ and $z$-motion for the same data set as in Fig.2c). The dashed-lines with associated colours are fitting of environmental damping being linearly proportional pressure. The lowest damping rate obtained is 2 mHz at $6\times10^{-6}$ mbar. The linear fit includes the pressure dependency only, while other effects may play a role at low pressure.}
    \label{fig2}
\end{figure*}

To achieve parametric feedback, unlike previous experiments~\cite{Li2011, PhysRevLett.109.103603} we lock an external oscillator to the particle's motion by means of a phase-locked loop (PLL) as also done in~\cite{Jain2016}. For a system with a high mechanical quality factor but with significant noise by damping factor $\Gamma_0$, this permits an accurate tracking of the motional phase of the mechanical harmonic oscillator, and allows a sinosoidal feedback to be applied relative to the noise induced phase jumps. In more detail, our system is driven by white noise, yet it can respond only to frequencies that fall within its bandwidth. Our capabilities of tracking the individual trajectories of the trapped nanoparticle as it undergoes collisions with the surrounding gas opens up the possibility for the assessment of the stochastic thermodynamics of the system. Therefore the phase of the mechanical oscillator can be extracted from the externally locked oscillator, and we implement parametric feedback with a controlled phase shift relative to the particle motion, in close analogy with early work on microcantilevers~\cite{Rugar91}. The trapped particle, which is driven strongly out of equilibrium by our driving and feedback mechanism embodies a very promising candidate for the study of dynamical processes and (quantum) fluctuation theorems~\cite{campisi2011, PhysRevLett.113.140601, PhysRevLett.115.190601}, while classical studies have been done already~\cite{Gieseler2013}. Such a study could turn out to be very informative for the characterisation of the properties of the environment into which the particle is effectively embedded~\cite{Millen2014n}.

The corresponding form of the power spectral density (PSD) of the oscillation is then straightforward to calculate from Eq.(\ref{eq:particlemotion}), leading to the modified Lorentzian response 
\begin{equation}\label{eq:PSD}
 S_{x}(\omega)=\frac{k_{B}T_0}{\pi m}\frac{\Gamma_{0}} {([\omega_{0}+\delta\omega]^{2}-\omega^2)^2+\omega^{2}[\Gamma_{0}+\delta\Gamma ]^{2}},
\end{equation}
with $k_B$ the Boltzmann constant. The interpretation of this result is then straightforward: by tuning the modulation depth of the trap and the relative phase between feedback signal and particle's motion, we are able to change the sign and amplitude of both $\delta\Gamma$ and $\delta\omega$, thus tuning the effective dynamics of the particle from an underdamped motion to an overdamped one, and blue or red-detuning its oscillation frequency with respect to that of the trap. This demonstrates the full control  that we have managed to achieve over the motion of the nanoparticle.

Assuming equilibration at the steady state of the motion, the equipartition theorem provides an effective temperature of the centre-of-mass motion given by $ T_{cm} = m(\omega_{0}+ \delta\omega)^{2}\langle x^{2}(t)\rangle / k_{B} $. As $ \delta\omega \ll \omega_{0} $ for all the experimental conditions of our work, we can write, $T_{cm}= T_0\Gamma_{0}/( \Gamma _{0}+ \delta \Gamma)$, where $T_0$ is the equilibrium temperature in the absence of parametric feedback cooling, which is here assumed to be 300K. We have experimentally checked the motion of the particle for different trapping laser powers and observed no dependency on such parameter. We instead observe a constant motional temperature, which verifies our assumption that the motion of the particle without active feedback is thermalised at room temperature.  We use the equation for $T_{cm}$ to evaluate the centre of mass temperature from fitting the Lorentzian Eq.(\ref{eq:PSD})  to the power spectral densities (see Supplement S1 and S7 for more details). 

\begin{figure*}
    \centering
    \includegraphics[width=0.9\textwidth]{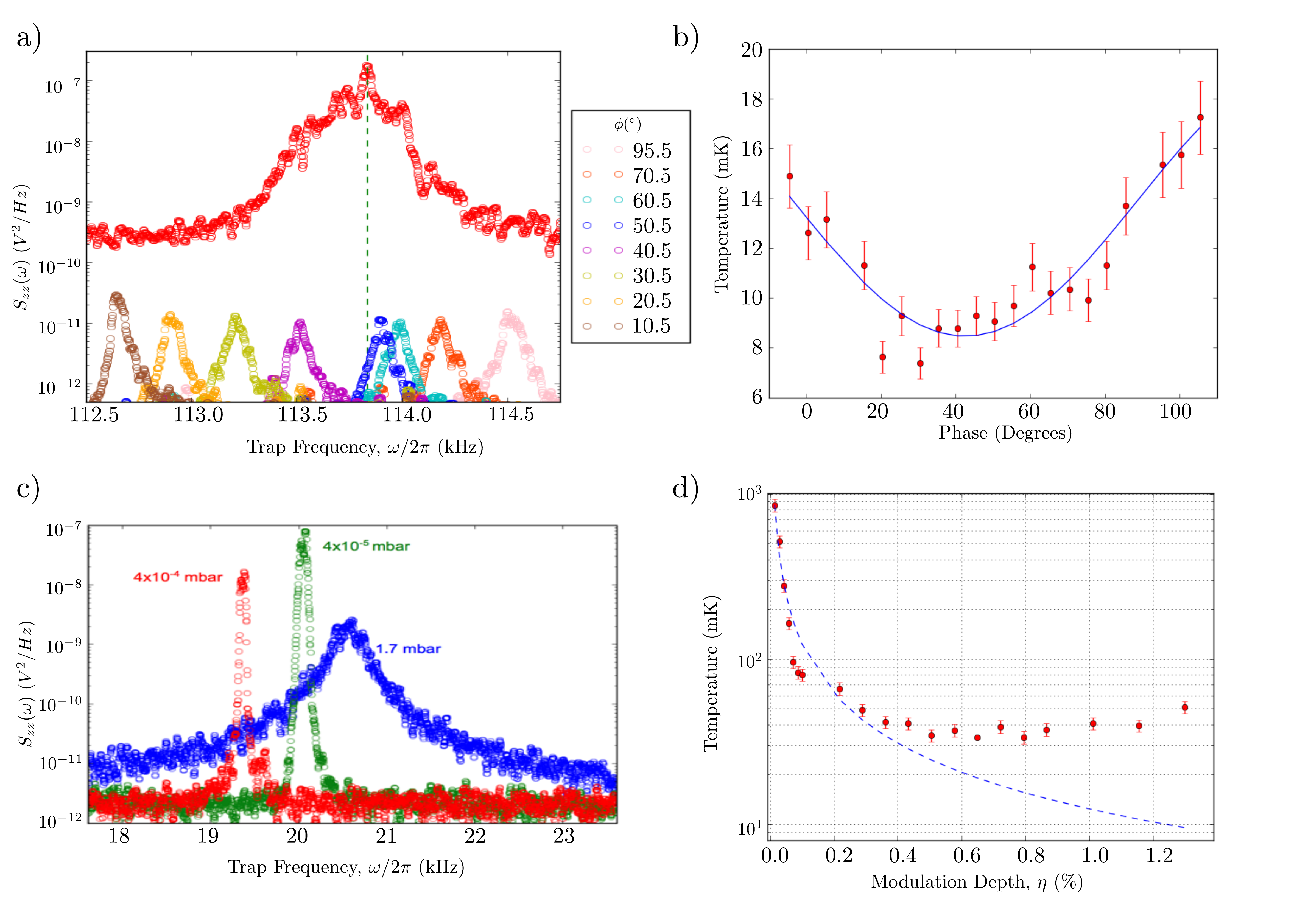}
    \caption{\textbf{Feedback dependent effects.} All data shown are for the $z$ motion of the particle.
    a) Feedback phase scan for a 100nm diameter particle at 10$^{-5}$mbar with different phase $\phi$ for feedback. PSD showing the dependence of the mechanical oscillator frequency $\omega_0$ when driven with a feedback signal of differing phases. The red data represents data for a uncooled particle and its central frequency denoted by a dashed line. 
    b) Extracted temperature from PSDs in a) versus feedback phase $\phi$, according to Eq.(\ref{phasemod}) and a fixed modulation depth $\eta$=0.44.
    c) PSD of a 100nm diameter particle at different pressure without feedback. Particles could be trapped for some hours without significant changes to the PSD. For pressure below 5$\times$10$^{-6}$mbar the particle was always lost from the trap if no feedback was applied.
    d) We show the cooling for different feedback modulation depths $\eta$ for a 100nm diameter particle at 1$\times$10$^{-6}$mbar, fitted with the Eq.(\ref{phasemod}). All other parameters have been kept constant and the feedback loop phase optimized for maximum cooling.}
    \label{fig3}
\end{figure*}

{\bf Discussion:}  Particles can be trapped at atmospheric pressure and kept optically trapped without feedback down to pressures on the order of 10$^{-5}$mbar, see Fig.~\ref{fig3}c). This indicates that using 1550nm trapping light and particles with presumably low absorption cross section minimises the effects of internal heating and ionisation, which would ultimately result in trap loss. With feedback the pressure could be reduced to the limits of the present setup. We trap particles of diameter ranging from 26nm to 160nm at comparable trapping laser powers.

We evaluate the radius $r$ and mass $m$ of the particle from fitting Eq.(\ref{eq:PSD}) to PSDs. For typical data we get $m$=$3.3(\pm0.7)\times10^{-19}$kg and $r=32(\pm6)$nm (see supplement S1 and S3 for details).The large error bars result from a large uncertainty in the measurement of the pressure in the vacuum chamber. To circumvent such limitations and to perform a parameter independent measurement we perform a further experiment to directly measure the amplitude of particle motion by scanning the wavelength of the trapping laser to modulate the relative amplitudes of the first and second harmonic PSD peaks of the z-motion of a trapped particle (see supplementary section S2 for more details). By comparing the change in amplitude between the first and second harmonic we are able to measure the amplitude of particle motion within the trap to be $119(\pm10)$nm. From the equipartition theorem: $m= k_{b}T_{cm}/\omega_{0}^{2}z_{0}^{2} $, with $T_{cm} = 300$K we can obtain a pressure independent measure of the particles mass $m=3(\pm0.5)\times10^{-19}$kg and thus a particle radius $r$=30$(\pm2)$nm. We evaluate the position resolution to be $S_{x, exp}=200$fm$/\sqrt{\text{Hz}}(\pm 8\%)$ for the same 60nm diameter particle as used for the wavelength scan. This position resolution is sufficient to measure a zero-point motion $\Delta x = 2.6$pm with an integration time of one second. 

The feedback compensates the thermalisation effect by gas collisions and stabilises the motion at a lower temperature. However, the minimum temperature that can be reached is dependent on the pressure of the system as shown in Fig.~\ref{fig2}c). This dependence is because the pressure damping from gas collisions acts as the dominant heating mechanism in our system. The motion of the particle can be cooled and the lowest temperatures $T_{cm}$ achieved are on the order of 1mK for x,y and z simultaneously at the lowest achieved pressure in the vacuum chamber of 1$\times10^{-6}$mbar.  The temperature reached is limited by both the pressure and the overall noise floor present in the experiment. An extrapolation of this pressure dependence indicates that motional ground state can be achieved at a pressure on the order of $10^{-9}$mbar, which is technically feasible although not implemented in our current setup.

The collisional damping rate $\Gamma_0$ is analysed for different pressures in Fig.~\ref{fig2}d), and found to be dominating the width of the PSD $\Delta\omega$ and used to analyse the mechanical factor as $Q_m = \omega_0 / \Delta\omega$. $Q_m$ is expected to be exceptionally high for the motion of a trapped particle \cite{Gieseler2013}, assuming the relation, $Q_m=\omega_0 / \Gamma_0$, to hold and a linear decrease of $\Gamma_0$ with decreasing pressure. Experimentally we find $Q_m=$4.4$\times$10$^7$ from the respective collisional damping data at 6$\times$10$^{-6}$mbar, where the damping rate is measured to be $\Gamma_0 /(2\pi)$=2mHz. Assuming that this scaling can be extrapolated to ultra-high vacuum, we find a maximum $Q_m$=10$^{12}$ at $10^{-9}$mbar, in agreement with previous studies~\cite{Gieseler2013}.

Feedback cooling rates achieved are on the order of 100Hz depending on the feedback modulation depth $\eta$ and phase $\phi$. Parametric feedback is by definition the modulation of the optical spring constant $k_{trap}$ at twice the trap frequency. Here we report on an effect of the phase of the feedback on the temperature and frequency of the motion of the trapped particle, see Figs.~\ref{fig3}a) and \ref{fig3}b). To study this effect in more detail we systematically change  $\phi$  to switch between parametric cooling and parametric heating. We find that starting from the optimal cooling situation, that there is heating in both directions when varying the phase, see Fig. \ref{fig3}a). The relationship between the feedback phase $\phi$, modulation depth $\eta$ and motional temperature $T_{cm}$ can be derived from a fluctuation theorem for a linear feedback modulation (see~\cite{GieselerNJP2015}) to be
\begin{equation}\label{phasemod}
T_{cm} = \frac{T_{0}}{1-\frac{\eta \omega_{0} \sin(2 \phi)}{2 \Gamma_{0}}} .
\end{equation}
We find good agreement between theory and experiment for the modulation of the feedback phase $\phi$, as shown in Fig.~\ref{fig3}b) where Eq.(\ref{phasemod}) is fitted to the experimental data with $\eta =$0.44. Increasing of the feedback modulation depth $\eta$ shows a decrease of the motional temperature in the experiment, which is shown fitted with Eq.(\ref{phasemod}) to the experimental data in Fig. \ref{fig3}d) and is in perfect agreement for modulations of $\eta <0.5\%$. However at larger modulation depths we find the effect of parametric feedback to cause heating instead.
 
We have demonstrated a parabolic mirror geometry for optically trapping of nanoparticles in high vacuum and achieve parametric feedback cooling in all three motional directions to a few mK. We find a trap frequency modulation effect, which depends on the phase of the parametric feedback.We measure the sensitivity of the system to be sufficient to measure a phonon equivalent displacement or zero-point motion. We find that the system is limited by detector noise and the pressures considered. We conclude that the implementation of ultra-high vacuum and reduction of system noise will pave the way to reaching a low phonon state of levitated optomechanics in this system. We find a single photon recoil limit for our experimental settings of $n=24$ phonons as the predicted limit towards ground state cooling (see supplement S8 for more details). Further studies with our setup at ultra-high vacuum conditions will show if this limit can be reached.

We demonstrate {\it feedback cooled nanoparticles} that can be utilized as a point-like source for matterwave interferometers~\cite{bateman2014near, romero2011large, scala2013matter, Wan2016}. Trapped nanoparticles hold promise for testing so-called collapse models~\cite{bassi2013models,PhysRevLettVinante} also by non-interferometric means~\cite{bahrami2014proposal, nimmrichter2014optomechanical, diosi2015testing, Goldwater2016}. The simple, robust and easily vacuum compatible mirror trap might provide useful technology for spectroscopy investigation of environmental independent nanoparticles properties, experiments in space~\cite{kaltenbaek2016macroscopic}, and the detection of speculative particles~\cite{bateman2015existence,PhysRevDRiedel}. The high $Q_m$ value and the precise position detection of 200fm/$\sqrt{\text{Hz}}$ for a single particle of mass in excess of atomic scales holds promise for various sensing application based on classical levitated optomechanics~\cite{PhysRevLettGeraci}.  Further experiments with levitated optomechanics will aim to demonstrate quantum control~\cite{PhysRevAYin} and to realise non-classical motional states of levitated optomechanics by continuous weak measurement~\cite{1367-2630-17-7-073019,PhysRevLett.70.548,wiseman2010quantum}.

{\bf Methods:} The optical trap is generated with light from a stabilised fibre laser. The output of the laser is intensity modulated by a AOM according to a feedback signal. The light then seeds an erbium doped fibre amplifier. The light is focused by a high numerical aperture (NA=0.995) parabolic aluminium mirror which is mounted in a vacuum chamber. We trap silica nanoparticles at vacuum as low as 1$\times$10$^{-6}$mbar and at trap frequency $\omega_0$. This trap frequency can be controlled by varying the laser power and hence the optical spring constant to achieve a $\omega_0$ in the range 10kHz to 300kHz. The particle is a driven dipole and hence radiates a field, a backwards propagating fraction of which is collected and collimated by the same parabolic mirror. Interference between this back scattered field $E_{scat}$, and the highly divergent field which passes without interacting with the particle through the focus  $E_{div}$ (the local oscillator reference field), provides interferometric position resolution. By allowing the reference field to diverge, we make the reference field amplitude comparable with that scattered by the particle, giving a large modulation visibility at the detector.

{\it Acknowledgements:--} We thank for stimulating discussions all participants of the OSA Incubator meeting on Levitated Optomechanics. We thank Maximilian Bazzi and Catalina Curceanu for providing low-noise electronics for a quadrant photo diode detector from which one channel was used for some of the experiments reported here. We further thank Nathan Cooper and Christopher Dawson for help with an earlier version of the mirror trap experiment. We thank Phil Connell and Gareth Savage for expert technical help during the realisation of the experimental setup. We wish to thank the UK funding agency EPSRC for support under grant (EP/J014664/1), the Foundational Questions Institute (FQXi), and the John F Templeton Foundation under grant (39530). MP thanks the EU FP7-funded Collaborative  Project TherMiQ (grant agreement 618074), the John Templeton Foundation (grant number 43467). We are grateful to Jelena Trbovic and Marco Brunner of Zurich Instruments for the extended loan of one of the precious lock-in amplifiers and Prof Harvey Rutt for lending to us the additional lock-in hardware.

\bibliographystyle{apsrev4-1}
\bibliography{references}

\begin{thebibliography}{46}%
\makeatletter
\providecommand \@ifxundefined [1]{%
 \@ifx{#1\undefined}
}%
\providecommand \@ifnum [1]{%
 \ifnum #1\expandafter \@firstoftwo
 \else \expandafter \@secondoftwo
 \fi
}%
\providecommand \@ifx [1]{%
 \ifx #1\expandafter \@firstoftwo
 \else \expandafter \@secondoftwo
 \fi
}%
\providecommand \natexlab [1]{#1}%
\providecommand \enquote  [1]{``#1''}%
\providecommand \bibnamefont  [1]{#1}%
\providecommand \bibfnamefont [1]{#1}%
\providecommand \citenamefont [1]{#1}%
\providecommand \href@noop [0]{\@secondoftwo}%
\providecommand \href [0]{\begingroup \@sanitize@url \@href}%
\providecommand \@href[1]{\@@startlink{#1}\@@href}%
\providecommand \@@href[1]{\endgroup#1\@@endlink}%
\providecommand \@sanitize@url [0]{\catcode `\\12\catcode `\$12\catcode
  `\&12\catcode `\#12\catcode `\^12\catcode `\_12\catcode `\%12\relax}%
\providecommand \@@startlink[1]{}%
\providecommand \@@endlink[0]{}%
\providecommand \url  [0]{\begingroup\@sanitize@url \@url }%
\providecommand \@url [1]{\endgroup\@href {#1}{\urlprefix }}%
\providecommand \urlprefix  [0]{URL }%
\providecommand \Eprint [0]{\href }%
\providecommand \doibase [0]{http://dx.doi.org/}%
\providecommand \selectlanguage [0]{\@gobble}%
\providecommand \bibinfo  [0]{\@secondoftwo}%
\providecommand \bibfield  [0]{\@secondoftwo}%
\providecommand \translation [1]{[#1]}%
\providecommand \BibitemOpen [0]{}%
\providecommand \bibitemStop [0]{}%
\providecommand \bibitemNoStop [0]{.\EOS\space}%
\providecommand \EOS [0]{\spacefactor3000\relax}%
\providecommand \BibitemShut  [1]{\csname bibitem#1\endcsname}%
\let\auto@bib@innerbib\@empty
\bibitem [{\citenamefont {Aspelmeyer}\ \emph {et~al.}(2014)\citenamefont
  {Aspelmeyer}, \citenamefont {Kippenberg},\ and\ \citenamefont
  {Marquardt}}]{RevModPhysCavityoptomechanics}%
  \BibitemOpen
  \bibfield  {author} {\bibinfo {author} {\bibfnamefont {M.}~\bibnamefont
  {Aspelmeyer}}, \bibinfo {author} {\bibfnamefont {T.~J.}\ \bibnamefont
  {Kippenberg}}, \ and\ \bibinfo {author} {\bibfnamefont {F.}~\bibnamefont
  {Marquardt}},\ }\href {\doibase 10.1103/RevModPhys.86.1391} {\bibfield
  {journal} {\bibinfo  {journal} {Rev. Mod. Phys.}\ }\textbf {\bibinfo {volume}
  {86}},\ \bibinfo {pages} {1391} (\bibinfo {year} {2014})}\BibitemShut
  {NoStop}%
\bibitem [{\citenamefont {Neukirch}\ \emph {et~al.}(2015)\citenamefont
  {Neukirch}, \citenamefont {Haartman}, \citenamefont {M},\ and\ \citenamefont
  {Vamivakas}}]{Levispin}%
  \BibitemOpen
  \bibfield  {author} {\bibinfo {author} {\bibfnamefont {L.~P.}\ \bibnamefont
  {Neukirch}}, \bibinfo {author} {\bibfnamefont {E.~v.}\ \bibnamefont
  {Haartman}}, \bibinfo {author} {\bibfnamefont {R.~J.}\ \bibnamefont {M}}, \
  and\ \bibinfo {author} {\bibfnamefont {A.~N.}\ \bibnamefont {Vamivakas}},\
  }\href {http://dx.doi.org/10.1038/nphoton.2015.162} {\bibfield  {journal}
  {\bibinfo  {journal} {Nat. Photon.}\ }\textbf {\bibinfo {volume} {9}},\
  \bibinfo {pages} {653–657} (\bibinfo {year} {2015})}\BibitemShut {NoStop}%
\bibitem [{\citenamefont {Arndt}\ and\ \citenamefont
  {Hornberger}(2014)}]{macro}%
  \BibitemOpen
  \bibfield  {author} {\bibinfo {author} {\bibfnamefont {M.}~\bibnamefont
  {Arndt}}\ and\ \bibinfo {author} {\bibfnamefont {K.}~\bibnamefont
  {Hornberger}},\ }\href {\doibase 10.1038/nphys2863} {\bibfield  {journal}
  {\bibinfo  {journal} {Nat Phys}\ }\textbf {\bibinfo {volume} {10}},\ \bibinfo
  {pages} {271} (\bibinfo {year} {2014})}\BibitemShut {NoStop}%
\bibitem [{\citenamefont {Bassi}\ \emph
  {et~al.}(2013{\natexlab{a}})\citenamefont {Bassi}, \citenamefont {Lochan},
  \citenamefont {Satin}, \citenamefont {Singh},\ and\ \citenamefont
  {Ulbricht}}]{RevModPhysBassi}%
  \BibitemOpen
  \bibfield  {author} {\bibinfo {author} {\bibfnamefont {A.}~\bibnamefont
  {Bassi}}, \bibinfo {author} {\bibfnamefont {K.}~\bibnamefont {Lochan}},
  \bibinfo {author} {\bibfnamefont {S.}~\bibnamefont {Satin}}, \bibinfo
  {author} {\bibfnamefont {T.~P.}\ \bibnamefont {Singh}}, \ and\ \bibinfo
  {author} {\bibfnamefont {H.}~\bibnamefont {Ulbricht}},\ }\href {\doibase
  10.1103/RevModPhys.85.471} {\bibfield  {journal} {\bibinfo  {journal} {Rev.
  Mod. Phys.}\ }\textbf {\bibinfo {volume} {85}},\ \bibinfo {pages} {471}
  (\bibinfo {year} {2013}{\natexlab{a}})}\BibitemShut {NoStop}%
\bibitem [{\citenamefont {Diosi}(1984)}]{Gravitation}%
  \BibitemOpen
  \bibfield  {author} {\bibinfo {author} {\bibfnamefont {L.}~\bibnamefont
  {Diosi}},\ }\href {\doibase http://dx.doi.org/10.1016/0375-9601(84)90397-9}
  {\bibfield  {journal} {\bibinfo  {journal} {Phys. Lett. A}\ }\textbf
  {\bibinfo {volume} {105}},\ \bibinfo {pages} {199 } (\bibinfo {year}
  {1984})}\BibitemShut {NoStop}%
\bibitem [{\citenamefont {Penrose}(1996)}]{Penrose1996581}%
  \BibitemOpen
  \bibfield  {author} {\bibinfo {author} {\bibfnamefont {R.}~\bibnamefont
  {Penrose}},\ }\href@noop {} {\bibfield  {journal} {\bibinfo  {journal} {Gen.
  Rel. Grav.}\ }\textbf {\bibinfo {volume} {28}},\ \bibinfo {pages} {581}
  (\bibinfo {year} {1996})}\BibitemShut {NoStop}%
\bibitem [{\citenamefont {Anastopoulos}\ and\ \citenamefont
  {Hu}(2015)}]{ISI000358741800023}%
  \BibitemOpen
  \bibfield  {author} {\bibinfo {author} {\bibfnamefont {C.}~\bibnamefont
  {Anastopoulos}}\ and\ \bibinfo {author} {\bibfnamefont {B.~L.}\ \bibnamefont
  {Hu}},\ }\href@noop {} {\bibfield  {journal} {\bibinfo  {journal} {Class.
  Quantum Grav.}\ }\textbf {\bibinfo {volume} {32}},\ \bibinfo {pages} {165022}
  (\bibinfo {year} {2015})}\BibitemShut {NoStop}%
\bibitem [{\citenamefont {Li}\ \emph {et~al.}(2011)\citenamefont {Li},
  \citenamefont {Kheifets},\ and\ \citenamefont {Raizen}}]{Li2011}%
  \BibitemOpen
  \bibfield  {author} {\bibinfo {author} {\bibfnamefont {T.}~\bibnamefont
  {Li}}, \bibinfo {author} {\bibfnamefont {S.}~\bibnamefont {Kheifets}}, \ and\
  \bibinfo {author} {\bibfnamefont {M.~G.}\ \bibnamefont {Raizen}},\ }\href
  {http://dx.doi.org/10.1038/nphys1952} {\bibfield  {journal} {\bibinfo
  {journal} {Nat. Phys.}\ }\textbf {\bibinfo {volume} {7}},\ \bibinfo {pages}
  {527} (\bibinfo {year} {2011})}\BibitemShut {NoStop}%
\bibitem [{\citenamefont {Gieseler}\ \emph {et~al.}(2012)\citenamefont
  {Gieseler}, \citenamefont {Deutsch}, \citenamefont {Quidant},\ and\
  \citenamefont {Novotny}}]{PhysRevLett.109.103603}%
  \BibitemOpen
  \bibfield  {author} {\bibinfo {author} {\bibfnamefont {J.}~\bibnamefont
  {Gieseler}}, \bibinfo {author} {\bibfnamefont {B.}~\bibnamefont {Deutsch}},
  \bibinfo {author} {\bibfnamefont {R.}~\bibnamefont {Quidant}}, \ and\
  \bibinfo {author} {\bibfnamefont {L.}~\bibnamefont {Novotny}},\ }\href
  {\doibase 10.1103/PhysRevLett.109.103603} {\bibfield  {journal} {\bibinfo
  {journal} {Phys. Rev. Lett.}\ }\textbf {\bibinfo {volume} {109}},\ \bibinfo
  {pages} {103603} (\bibinfo {year} {2012})}\BibitemShut {NoStop}%
\bibitem [{\citenamefont {Jain}\ \emph {et~al.}(2016)\citenamefont {Jain},
  \citenamefont {Gieseler}, \citenamefont {Moritz}, \citenamefont {Dellago},
  \citenamefont {Quidant},\ and\ \citenamefont {Novotny}}]{Jain2016}%
  \BibitemOpen
  \bibfield  {author} {\bibinfo {author} {\bibfnamefont {V.}~\bibnamefont
  {Jain}}, \bibinfo {author} {\bibfnamefont {J.}~\bibnamefont {Gieseler}},
  \bibinfo {author} {\bibfnamefont {C.}~\bibnamefont {Moritz}}, \bibinfo
  {author} {\bibfnamefont {C.}~\bibnamefont {Dellago}}, \bibinfo {author}
  {\bibfnamefont {R.}~\bibnamefont {Quidant}}, \ and\ \bibinfo {author}
  {\bibfnamefont {L.}~\bibnamefont {Novotny}},\ }\href {\doibase
  10.1103/PhysRevLett.116.243601} {\bibfield  {journal} {\bibinfo  {journal}
  {Phys. Rev. Lett.}\ }\textbf {\bibinfo {volume} {116}},\ \bibinfo {pages}
  {243601} (\bibinfo {year} {2016})}\BibitemShut {NoStop}%
\bibitem [{\citenamefont {Asenbaum}\ \emph {et~al.}(2013)\citenamefont
  {Asenbaum}, \citenamefont {Kuhn}, \citenamefont {Nimmrichter}, \citenamefont
  {Sezer},\ and\ \citenamefont {Arndt}}]{Asenbaum2013}%
  \BibitemOpen
  \bibfield  {author} {\bibinfo {author} {\bibfnamefont {P.}~\bibnamefont
  {Asenbaum}}, \bibinfo {author} {\bibfnamefont {S.}~\bibnamefont {Kuhn}},
  \bibinfo {author} {\bibfnamefont {S.}~\bibnamefont {Nimmrichter}}, \bibinfo
  {author} {\bibfnamefont {U.}~\bibnamefont {Sezer}}, \ and\ \bibinfo {author}
  {\bibfnamefont {M.}~\bibnamefont {Arndt}},\ }\href@noop {} {\bibfield
  {journal} {\bibinfo  {journal} {Nat. Commun.}\ }\textbf {\bibinfo {volume}
  {4}},\ \bibinfo {pages} {3743} (\bibinfo {year} {2013})}\BibitemShut
  {NoStop}%
\bibitem [{\citenamefont {Kiesel}\ \emph {et~al.}(2013)\citenamefont {Kiesel},
  \citenamefont {Blaser}, \citenamefont {Deli{\'c}}, \citenamefont {Grass},
  \citenamefont {Kaltenbaek},\ and\ \citenamefont
  {Aspelmeyer}}]{kiesel2013cavity}%
  \BibitemOpen
  \bibfield  {author} {\bibinfo {author} {\bibfnamefont {N.}~\bibnamefont
  {Kiesel}}, \bibinfo {author} {\bibfnamefont {F.}~\bibnamefont {Blaser}},
  \bibinfo {author} {\bibfnamefont {U.}~\bibnamefont {Deli{\'c}}}, \bibinfo
  {author} {\bibfnamefont {D.}~\bibnamefont {Grass}}, \bibinfo {author}
  {\bibfnamefont {R.}~\bibnamefont {Kaltenbaek}}, \ and\ \bibinfo {author}
  {\bibfnamefont {M.}~\bibnamefont {Aspelmeyer}},\ }\href@noop {} {\bibfield
  {journal} {\bibinfo  {journal} {PNAS}\ }\textbf {\bibinfo {volume} {110}},\
  \bibinfo {pages} {14180} (\bibinfo {year} {2013})}\BibitemShut {NoStop}%
\bibitem [{\citenamefont {Chang}\ \emph {et~al.}(2010)\citenamefont {Chang},
  \citenamefont {Regal}, \citenamefont {Papp}, \citenamefont {Wilson},
  \citenamefont {Ye}, \citenamefont {Painter}, \citenamefont {Kimble},\ and\
  \citenamefont {Zoller}}]{Chang19012010}%
  \BibitemOpen
  \bibfield  {author} {\bibinfo {author} {\bibfnamefont {D.~E.}\ \bibnamefont
  {Chang}}, \bibinfo {author} {\bibfnamefont {C.~A.}\ \bibnamefont {Regal}},
  \bibinfo {author} {\bibfnamefont {S.~B.}\ \bibnamefont {Papp}}, \bibinfo
  {author} {\bibfnamefont {D.~J.}\ \bibnamefont {Wilson}}, \bibinfo {author}
  {\bibfnamefont {J.}~\bibnamefont {Ye}}, \bibinfo {author} {\bibfnamefont
  {O.}~\bibnamefont {Painter}}, \bibinfo {author} {\bibfnamefont {H.~J.}\
  \bibnamefont {Kimble}}, \ and\ \bibinfo {author} {\bibfnamefont
  {P.}~\bibnamefont {Zoller}},\ }\href@noop {} {\bibfield  {journal} {\bibinfo
  {journal} {PNAS}\ }\textbf {\bibinfo {volume} {107}},\ \bibinfo {pages}
  {1005} (\bibinfo {year} {2010})}\BibitemShut {NoStop}%
\bibitem [{\citenamefont {Millen}\ \emph {et~al.}(2015)\citenamefont {Millen},
  \citenamefont {Fonseca}, \citenamefont {Mavrogordatos}, \citenamefont
  {Monteiro},\ and\ \citenamefont {Barker}}]{millen2015cavity}%
  \BibitemOpen
  \bibfield  {author} {\bibinfo {author} {\bibfnamefont {J.}~\bibnamefont
  {Millen}}, \bibinfo {author} {\bibfnamefont {P.}~\bibnamefont {Fonseca}},
  \bibinfo {author} {\bibfnamefont {T.}~\bibnamefont {Mavrogordatos}}, \bibinfo
  {author} {\bibfnamefont {T.}~\bibnamefont {Monteiro}}, \ and\ \bibinfo
  {author} {\bibfnamefont {P.}~\bibnamefont {Barker}},\ }\href@noop {}
  {\bibfield  {journal} {\bibinfo  {journal} {Phys. Rev. Lett.}\ }\textbf
  {\bibinfo {volume} {114}},\ \bibinfo {pages} {123602} (\bibinfo {year}
  {2015})}\BibitemShut {NoStop}%
\bibitem [{\citenamefont {Arita}\ \emph {et~al.}(2013)\citenamefont {Arita},
  \citenamefont {Mazilu},\ and\ \citenamefont {Dholakia}}]{arita2013laser}%
  \BibitemOpen
  \bibfield  {author} {\bibinfo {author} {\bibfnamefont {Y.}~\bibnamefont
  {Arita}}, \bibinfo {author} {\bibfnamefont {M.}~\bibnamefont {Mazilu}}, \
  and\ \bibinfo {author} {\bibfnamefont {K.}~\bibnamefont {Dholakia}},\
  }\href@noop {} {\bibfield  {journal} {\bibinfo  {journal} {Nat. Commun.}\
  }\textbf {\bibinfo {volume} {4}},\ \bibinfo {pages} {2374} (\bibinfo {year}
  {2013})}\BibitemShut {NoStop}%
\bibitem [{\citenamefont {Kuhn}\ \emph {et~al.}(2015)\citenamefont {Kuhn},
  \citenamefont {Asenbaum}, \citenamefont {Kosloff}, \citenamefont {Sclafani},
  \citenamefont {Stickler}, \citenamefont {Nimmrichter}, \citenamefont
  {Hornberger}, \citenamefont {Cheshnovsky}, \citenamefont {Patolsky},\ and\
  \citenamefont {Arndt}}]{kuhn2015cavity}%
  \BibitemOpen
  \bibfield  {author} {\bibinfo {author} {\bibfnamefont {S.}~\bibnamefont
  {Kuhn}}, \bibinfo {author} {\bibfnamefont {P.}~\bibnamefont {Asenbaum}},
  \bibinfo {author} {\bibfnamefont {A.}~\bibnamefont {Kosloff}}, \bibinfo
  {author} {\bibfnamefont {M.}~\bibnamefont {Sclafani}}, \bibinfo {author}
  {\bibfnamefont {B.~A.}\ \bibnamefont {Stickler}}, \bibinfo {author}
  {\bibfnamefont {S.}~\bibnamefont {Nimmrichter}}, \bibinfo {author}
  {\bibfnamefont {K.}~\bibnamefont {Hornberger}}, \bibinfo {author}
  {\bibfnamefont {O.}~\bibnamefont {Cheshnovsky}}, \bibinfo {author}
  {\bibfnamefont {F.}~\bibnamefont {Patolsky}}, \ and\ \bibinfo {author}
  {\bibfnamefont {M.}~\bibnamefont {Arndt}},\ }\href@noop {} {\bibfield
  {journal} {\bibinfo  {journal} {Nano Lett.}\ }\textbf {\bibinfo {volume}
  {15}},\ \bibinfo {pages} {5604} (\bibinfo {year} {2015})}\BibitemShut
  {NoStop}%
\bibitem [{\citenamefont {Hoang}\ \emph {et~al.}(2016)\citenamefont {Hoang},
  \citenamefont {Ma}, \citenamefont {Ahn}, \citenamefont {Bang}, \citenamefont
  {Robicheaux}, \citenamefont {Yin},\ and\ \citenamefont
  {Li}}]{PhysRevLett.117.123604}%
  \BibitemOpen
  \bibfield  {author} {\bibinfo {author} {\bibfnamefont {T.~M.}\ \bibnamefont
  {Hoang}}, \bibinfo {author} {\bibfnamefont {Y.}~\bibnamefont {Ma}}, \bibinfo
  {author} {\bibfnamefont {J.}~\bibnamefont {Ahn}}, \bibinfo {author}
  {\bibfnamefont {J.}~\bibnamefont {Bang}}, \bibinfo {author} {\bibfnamefont
  {F.}~\bibnamefont {Robicheaux}}, \bibinfo {author} {\bibfnamefont {Z.-Q.}\
  \bibnamefont {Yin}}, \ and\ \bibinfo {author} {\bibfnamefont
  {T.}~\bibnamefont {Li}},\ }\href {\doibase 10.1103/PhysRevLett.117.123604}
  {\bibfield  {journal} {\bibinfo  {journal} {Phys. Rev. Lett.}\ }\textbf
  {\bibinfo {volume} {117}},\ \bibinfo {pages} {123604} (\bibinfo {year}
  {2016})}\BibitemShut {NoStop}%
\bibitem [{\citenamefont {Rashid}\ \emph {et~al.}(2016)\citenamefont {Rashid},
  \citenamefont {Tufarelli}, \citenamefont {Bateman}, \citenamefont {Vovrosh},
  \citenamefont {Hempston}, \citenamefont {Kim},\ and\ \citenamefont
  {Ulbricht}}]{rashid2016experimental}%
  \BibitemOpen
  \bibfield  {author} {\bibinfo {author} {\bibfnamefont {M.}~\bibnamefont
  {Rashid}}, \bibinfo {author} {\bibfnamefont {T.}~\bibnamefont {Tufarelli}},
  \bibinfo {author} {\bibfnamefont {J.}~\bibnamefont {Bateman}}, \bibinfo
  {author} {\bibfnamefont {J.}~\bibnamefont {Vovrosh}}, \bibinfo {author}
  {\bibfnamefont {D.}~\bibnamefont {Hempston}}, \bibinfo {author}
  {\bibfnamefont {M.}~\bibnamefont {Kim}}, \ and\ \bibinfo {author}
  {\bibfnamefont {H.}~\bibnamefont {Ulbricht}},\ }\href@noop {} {\bibfield
  {journal} {\bibinfo  {journal} {arXiv preprint arXiv:1607.05509}\ } (\bibinfo
  {year} {2016})}\BibitemShut {NoStop}%
\bibitem [{\citenamefont {Gieseler}\ \emph {et~al.}(2013)\citenamefont
  {Gieseler}, \citenamefont {Novotny},\ and\ \citenamefont
  {Quidant}}]{Gieseler2013}%
  \BibitemOpen
  \bibfield  {author} {\bibinfo {author} {\bibfnamefont {J.}~\bibnamefont
  {Gieseler}}, \bibinfo {author} {\bibfnamefont {L.}~\bibnamefont {Novotny}}, \
  and\ \bibinfo {author} {\bibfnamefont {R.}~\bibnamefont {Quidant}},\
  }\href@noop {} {\bibfield  {journal} {\bibinfo  {journal} {Nat. Phys.}\
  }\textbf {\bibinfo {volume} {9}},\ \bibinfo {pages} {806} (\bibinfo {year}
  {2013})}\BibitemShut {NoStop}%
\bibitem [{\citenamefont {Rugar}\ and\ \citenamefont
  {Gr\"utter}(1991)}]{Rugar91}%
  \BibitemOpen
  \bibfield  {author} {\bibinfo {author} {\bibfnamefont {D.}~\bibnamefont
  {Rugar}}\ and\ \bibinfo {author} {\bibfnamefont {P.}~\bibnamefont
  {Gr\"utter}},\ }\href {\doibase 10.1103/PhysRevLett.67.699} {\bibfield
  {journal} {\bibinfo  {journal} {Phys. Rev. Lett.}\ }\textbf {\bibinfo
  {volume} {67}},\ \bibinfo {pages} {699} (\bibinfo {year} {1991})}\BibitemShut
  {NoStop}%
\bibitem [{\citenamefont {Campisi}\ \emph {et~al.}(2011)\citenamefont
  {Campisi}, \citenamefont {H{\"a}nggi},\ and\ \citenamefont
  {Talkner}}]{campisi2011}%
  \BibitemOpen
  \bibfield  {author} {\bibinfo {author} {\bibfnamefont {M.}~\bibnamefont
  {Campisi}}, \bibinfo {author} {\bibfnamefont {P.}~\bibnamefont {H{\"a}nggi}},
  \ and\ \bibinfo {author} {\bibfnamefont {P.}~\bibnamefont {Talkner}},\
  }\href@noop {} {\bibfield  {journal} {\bibinfo  {journal} {Reviews of Modern
  Physics}\ }\textbf {\bibinfo {volume} {83}},\ \bibinfo {pages} {771}
  (\bibinfo {year} {2011})}\BibitemShut {NoStop}%
\bibitem [{\citenamefont {Batalh\~ao}\ \emph {et~al.}(2014)\citenamefont
  {Batalh\~ao}, \citenamefont {Souza}, \citenamefont {Mazzola}, \citenamefont
  {Auccaise}, \citenamefont {Sarthour}, \citenamefont {Oliveira}, \citenamefont
  {Goold}, \citenamefont {De~Chiara}, \citenamefont {Paternostro},\ and\
  \citenamefont {Serra}}]{PhysRevLett.113.140601}%
  \BibitemOpen
  \bibfield  {author} {\bibinfo {author} {\bibfnamefont {T.~B.}\ \bibnamefont
  {Batalh\~ao}}, \bibinfo {author} {\bibfnamefont {A.~M.}\ \bibnamefont
  {Souza}}, \bibinfo {author} {\bibfnamefont {L.}~\bibnamefont {Mazzola}},
  \bibinfo {author} {\bibfnamefont {R.}~\bibnamefont {Auccaise}}, \bibinfo
  {author} {\bibfnamefont {R.~S.}\ \bibnamefont {Sarthour}}, \bibinfo {author}
  {\bibfnamefont {I.~S.}\ \bibnamefont {Oliveira}}, \bibinfo {author}
  {\bibfnamefont {J.}~\bibnamefont {Goold}}, \bibinfo {author} {\bibfnamefont
  {G.}~\bibnamefont {De~Chiara}}, \bibinfo {author} {\bibfnamefont
  {M.}~\bibnamefont {Paternostro}}, \ and\ \bibinfo {author} {\bibfnamefont
  {R.~M.}\ \bibnamefont {Serra}},\ }\href {\doibase
  10.1103/PhysRevLett.113.140601} {\bibfield  {journal} {\bibinfo  {journal}
  {Phys. Rev. Lett.}\ }\textbf {\bibinfo {volume} {113}},\ \bibinfo {pages}
  {140601} (\bibinfo {year} {2014})}\BibitemShut {NoStop}%
\bibitem [{\citenamefont {Batalh\~ao}\ \emph {et~al.}(2015)\citenamefont
  {Batalh\~ao}, \citenamefont {Souza}, \citenamefont {Sarthour}, \citenamefont
  {Oliveira}, \citenamefont {Paternostro}, \citenamefont {Lutz},\ and\
  \citenamefont {Serra}}]{PhysRevLett.115.190601}%
  \BibitemOpen
  \bibfield  {author} {\bibinfo {author} {\bibfnamefont {T.~B.}\ \bibnamefont
  {Batalh\~ao}}, \bibinfo {author} {\bibfnamefont {A.~M.}\ \bibnamefont
  {Souza}}, \bibinfo {author} {\bibfnamefont {R.~S.}\ \bibnamefont {Sarthour}},
  \bibinfo {author} {\bibfnamefont {I.~S.}\ \bibnamefont {Oliveira}}, \bibinfo
  {author} {\bibfnamefont {M.}~\bibnamefont {Paternostro}}, \bibinfo {author}
  {\bibfnamefont {E.}~\bibnamefont {Lutz}}, \ and\ \bibinfo {author}
  {\bibfnamefont {R.~M.}\ \bibnamefont {Serra}},\ }\href {\doibase
  10.1103/PhysRevLett.115.190601} {\bibfield  {journal} {\bibinfo  {journal}
  {Phys. Rev. Lett.}\ }\textbf {\bibinfo {volume} {115}},\ \bibinfo {pages}
  {190601} (\bibinfo {year} {2015})}\BibitemShut {NoStop}%
\bibitem [{\citenamefont {Millen}\ \emph {et~al.}(2014)\citenamefont {Millen},
  \citenamefont {Deesuwan}, \citenamefont {Barker},\ and\ \citenamefont
  {Anders}}]{Millen2014n}%
  \BibitemOpen
  \bibfield  {author} {\bibinfo {author} {\bibfnamefont {J.}~\bibnamefont
  {Millen}}, \bibinfo {author} {\bibfnamefont {T.}~\bibnamefont {Deesuwan}},
  \bibinfo {author} {\bibfnamefont {P.}~\bibnamefont {Barker}}, \ and\ \bibinfo
  {author} {\bibfnamefont {J.}~\bibnamefont {Anders}},\ }\href@noop {}
  {\bibfield  {journal} {\bibinfo  {journal} {Nat. Nano.}\ }\textbf {\bibinfo
  {volume} {9}},\ \bibinfo {pages} {425} (\bibinfo {year} {2014})}\BibitemShut
  {NoStop}%
\bibitem [{\citenamefont {Gieseler}\ \emph {et~al.}(2015)\citenamefont
  {Gieseler}, \citenamefont {Novotny}, \citenamefont {Moritz},\ and\
  \citenamefont {Dellago}}]{GieselerNJP2015}%
  \BibitemOpen
  \bibfield  {author} {\bibinfo {author} {\bibfnamefont {J.}~\bibnamefont
  {Gieseler}}, \bibinfo {author} {\bibfnamefont {L.}~\bibnamefont {Novotny}},
  \bibinfo {author} {\bibfnamefont {C.}~\bibnamefont {Moritz}}, \ and\ \bibinfo
  {author} {\bibfnamefont {C.}~\bibnamefont {Dellago}},\ }\href
  {http://stacks.iop.org/1367-2630/17/i=4/a=045011} {\bibfield  {journal}
  {\bibinfo  {journal} {New Journal of Physics}\ }\textbf {\bibinfo {volume}
  {17}},\ \bibinfo {pages} {045011} (\bibinfo {year} {2015})}\BibitemShut
  {NoStop}%
\bibitem [{\citenamefont {Bateman}\ \emph {et~al.}(2014)\citenamefont
  {Bateman}, \citenamefont {Nimmrichter}, \citenamefont {Hornberger},\ and\
  \citenamefont {Ulbricht}}]{bateman2014near}%
  \BibitemOpen
  \bibfield  {author} {\bibinfo {author} {\bibfnamefont {J.}~\bibnamefont
  {Bateman}}, \bibinfo {author} {\bibfnamefont {S.}~\bibnamefont
  {Nimmrichter}}, \bibinfo {author} {\bibfnamefont {K.}~\bibnamefont
  {Hornberger}}, \ and\ \bibinfo {author} {\bibfnamefont {H.}~\bibnamefont
  {Ulbricht}},\ }\href@noop {} {\bibfield  {journal} {\bibinfo  {journal} {Nat.
  Commun.}\ }\textbf {\bibinfo {volume} {5}},\ \bibinfo {pages} {4788}
  (\bibinfo {year} {2014})}\BibitemShut {NoStop}%
\bibitem [{\citenamefont {Romero-Isart}\ \emph {et~al.}(2011)\citenamefont
  {Romero-Isart}, \citenamefont {Pflanzer}, \citenamefont {Blaser},
  \citenamefont {Kaltenbaek}, \citenamefont {Kiesel}, \citenamefont
  {Aspelmeyer},\ and\ \citenamefont {Cirac}}]{romero2011large}%
  \BibitemOpen
  \bibfield  {author} {\bibinfo {author} {\bibfnamefont {O.}~\bibnamefont
  {Romero-Isart}}, \bibinfo {author} {\bibfnamefont {A.~C.}\ \bibnamefont
  {Pflanzer}}, \bibinfo {author} {\bibfnamefont {F.}~\bibnamefont {Blaser}},
  \bibinfo {author} {\bibfnamefont {R.}~\bibnamefont {Kaltenbaek}}, \bibinfo
  {author} {\bibfnamefont {N.}~\bibnamefont {Kiesel}}, \bibinfo {author}
  {\bibfnamefont {M.}~\bibnamefont {Aspelmeyer}}, \ and\ \bibinfo {author}
  {\bibfnamefont {J.~I.}\ \bibnamefont {Cirac}},\ }\href@noop {} {\bibfield
  {journal} {\bibinfo  {journal} {Phys. Rev. Lett.}\ }\textbf {\bibinfo
  {volume} {107}},\ \bibinfo {pages} {020405} (\bibinfo {year}
  {2011})}\BibitemShut {NoStop}%
\bibitem [{\citenamefont {Scala}\ \emph {et~al.}(2013)\citenamefont {Scala},
  \citenamefont {Kim}, \citenamefont {Morley}, \citenamefont {Barker},\ and\
  \citenamefont {Bose}}]{scala2013matter}%
  \BibitemOpen
  \bibfield  {author} {\bibinfo {author} {\bibfnamefont {M.}~\bibnamefont
  {Scala}}, \bibinfo {author} {\bibfnamefont {M.}~\bibnamefont {Kim}}, \bibinfo
  {author} {\bibfnamefont {G.}~\bibnamefont {Morley}}, \bibinfo {author}
  {\bibfnamefont {P.}~\bibnamefont {Barker}}, \ and\ \bibinfo {author}
  {\bibfnamefont {S.}~\bibnamefont {Bose}},\ }\href@noop {} {\bibfield
  {journal} {\bibinfo  {journal} {Phys. Rev. Lett.}\ }\textbf {\bibinfo
  {volume} {111}},\ \bibinfo {pages} {180403} (\bibinfo {year}
  {2013})}\BibitemShut {NoStop}%
\bibitem [{\citenamefont {Wan}\ \emph {et~al.}(2016)\citenamefont {Wan},
  \citenamefont {Scala}, \citenamefont {Morley}, \citenamefont {Rahman},
  \citenamefont {Ulbricht}, \citenamefont {Bateman}, \citenamefont {Barker},
  \citenamefont {Bose},\ and\ \citenamefont {Kim}}]{Wan2016}%
  \BibitemOpen
  \bibfield  {author} {\bibinfo {author} {\bibfnamefont {C.}~\bibnamefont
  {Wan}}, \bibinfo {author} {\bibfnamefont {M.}~\bibnamefont {Scala}}, \bibinfo
  {author} {\bibfnamefont {G.~W.}\ \bibnamefont {Morley}}, \bibinfo {author}
  {\bibfnamefont {A.~A.}\ \bibnamefont {Rahman}}, \bibinfo {author}
  {\bibfnamefont {H.}~\bibnamefont {Ulbricht}}, \bibinfo {author}
  {\bibfnamefont {J.}~\bibnamefont {Bateman}}, \bibinfo {author} {\bibfnamefont
  {P.~F.}\ \bibnamefont {Barker}}, \bibinfo {author} {\bibfnamefont
  {S.}~\bibnamefont {Bose}}, \ and\ \bibinfo {author} {\bibfnamefont {M.~S.}\
  \bibnamefont {Kim}},\ }\href {\doibase 10.1103/PhysRevLett.117.143003}
  {\bibfield  {journal} {\bibinfo  {journal} {Phys. Rev. Lett.}\ }\textbf
  {\bibinfo {volume} {117}},\ \bibinfo {pages} {143003} (\bibinfo {year}
  {2016})}\BibitemShut {NoStop}%
\bibitem [{\citenamefont {Bassi}\ \emph
  {et~al.}(2013{\natexlab{b}})\citenamefont {Bassi}, \citenamefont {Lochan},
  \citenamefont {Satin}, \citenamefont {Singh},\ and\ \citenamefont
  {Ulbricht}}]{bassi2013models}%
  \BibitemOpen
  \bibfield  {author} {\bibinfo {author} {\bibfnamefont {A.}~\bibnamefont
  {Bassi}}, \bibinfo {author} {\bibfnamefont {K.}~\bibnamefont {Lochan}},
  \bibinfo {author} {\bibfnamefont {S.}~\bibnamefont {Satin}}, \bibinfo
  {author} {\bibfnamefont {T.~P.}\ \bibnamefont {Singh}}, \ and\ \bibinfo
  {author} {\bibfnamefont {H.}~\bibnamefont {Ulbricht}},\ }\href@noop {}
  {\bibfield  {journal} {\bibinfo  {journal} {Reviews of Modern Physics}\
  }\textbf {\bibinfo {volume} {85}},\ \bibinfo {pages} {471} (\bibinfo {year}
  {2013}{\natexlab{b}})}\BibitemShut {NoStop}%
\bibitem [{\citenamefont {Vinante}\ \emph {et~al.}(2016)\citenamefont
  {Vinante}, \citenamefont {Bahrami}, \citenamefont {Bassi}, \citenamefont
  {Usenko}, \citenamefont {Wijts},\ and\ \citenamefont
  {Oosterkamp}}]{PhysRevLettVinante}%
  \BibitemOpen
  \bibfield  {author} {\bibinfo {author} {\bibfnamefont {A.}~\bibnamefont
  {Vinante}}, \bibinfo {author} {\bibfnamefont {M.}~\bibnamefont {Bahrami}},
  \bibinfo {author} {\bibfnamefont {A.}~\bibnamefont {Bassi}}, \bibinfo
  {author} {\bibfnamefont {O.}~\bibnamefont {Usenko}}, \bibinfo {author}
  {\bibfnamefont {G.}~\bibnamefont {Wijts}}, \ and\ \bibinfo {author}
  {\bibfnamefont {T.~H.}\ \bibnamefont {Oosterkamp}},\ }\href {\doibase
  10.1103/PhysRevLett.116.090402} {\bibfield  {journal} {\bibinfo  {journal}
  {Phys. Rev. Lett.}\ }\textbf {\bibinfo {volume} {116}},\ \bibinfo {pages}
  {090402} (\bibinfo {year} {2016})}\BibitemShut {NoStop}%
\bibitem [{\citenamefont {Bahrami}\ \emph {et~al.}(2014)\citenamefont
  {Bahrami}, \citenamefont {Paternostro}, \citenamefont {Bassi},\ and\
  \citenamefont {Ulbricht}}]{bahrami2014proposal}%
  \BibitemOpen
  \bibfield  {author} {\bibinfo {author} {\bibfnamefont {M.}~\bibnamefont
  {Bahrami}}, \bibinfo {author} {\bibfnamefont {M.}~\bibnamefont
  {Paternostro}}, \bibinfo {author} {\bibfnamefont {A.}~\bibnamefont {Bassi}},
  \ and\ \bibinfo {author} {\bibfnamefont {H.}~\bibnamefont {Ulbricht}},\
  }\href@noop {} {\bibfield  {journal} {\bibinfo  {journal} {Phys. Rev. Lett.}\
  }\textbf {\bibinfo {volume} {112}},\ \bibinfo {pages} {210404} (\bibinfo
  {year} {2014})}\BibitemShut {NoStop}%
\bibitem [{\citenamefont {Nimmrichter}\ \emph {et~al.}(2014)\citenamefont
  {Nimmrichter}, \citenamefont {Hornberger},\ and\ \citenamefont
  {Hammerer}}]{nimmrichter2014optomechanical}%
  \BibitemOpen
  \bibfield  {author} {\bibinfo {author} {\bibfnamefont {S.}~\bibnamefont
  {Nimmrichter}}, \bibinfo {author} {\bibfnamefont {K.}~\bibnamefont
  {Hornberger}}, \ and\ \bibinfo {author} {\bibfnamefont {K.}~\bibnamefont
  {Hammerer}},\ }\href@noop {} {\bibfield  {journal} {\bibinfo  {journal}
  {Phys. Rev. Lett.}\ }\textbf {\bibinfo {volume} {113}},\ \bibinfo {pages}
  {020405} (\bibinfo {year} {2014})}\BibitemShut {NoStop}%
\bibitem [{\citenamefont {Diosi}(2015)}]{diosi2015testing}%
  \BibitemOpen
  \bibfield  {author} {\bibinfo {author} {\bibfnamefont {L.}~\bibnamefont
  {Diosi}},\ }\href@noop {} {\bibfield  {journal} {\bibinfo  {journal} {Phys.
  Rev. Lett.}\ }\textbf {\bibinfo {volume} {114}},\ \bibinfo {pages} {050403}
  (\bibinfo {year} {2015})}\BibitemShut {NoStop}%
\bibitem [{\citenamefont {Goldwater}\ \emph {et~al.}(2016)\citenamefont
  {Goldwater}, \citenamefont {Paternostro},\ and\ \citenamefont
  {Barker}}]{Goldwater2016}%
  \BibitemOpen
  \bibfield  {author} {\bibinfo {author} {\bibfnamefont {D.}~\bibnamefont
  {Goldwater}}, \bibinfo {author} {\bibfnamefont {M.}~\bibnamefont
  {Paternostro}}, \ and\ \bibinfo {author} {\bibfnamefont {P.~F.}\ \bibnamefont
  {Barker}},\ }\href {\doibase 10.1103/PhysRevA.94.010104} {\bibfield
  {journal} {\bibinfo  {journal} {Phys. Rev. A}\ }\textbf {\bibinfo {volume}
  {94}},\ \bibinfo {pages} {010104} (\bibinfo {year} {2016})}\BibitemShut
  {NoStop}%
\bibitem [{\citenamefont {Kaltenbaek}\ \emph {et~al.}(2016)\citenamefont
  {Kaltenbaek}, \citenamefont {Aspelmeyer}, \citenamefont {Barker},
  \citenamefont {Bassi}, \citenamefont {Bateman}, \citenamefont {Bongs},
  \citenamefont {Bose}, \citenamefont {Braxmaier}, \citenamefont {Brukner},
  \citenamefont {Christophe} \emph {et~al.}}]{kaltenbaek2016macroscopic}%
  \BibitemOpen
  \bibfield  {author} {\bibinfo {author} {\bibfnamefont {R.}~\bibnamefont
  {Kaltenbaek}}, \bibinfo {author} {\bibfnamefont {M.}~\bibnamefont
  {Aspelmeyer}}, \bibinfo {author} {\bibfnamefont {P.~F.}\ \bibnamefont
  {Barker}}, \bibinfo {author} {\bibfnamefont {A.}~\bibnamefont {Bassi}},
  \bibinfo {author} {\bibfnamefont {J.}~\bibnamefont {Bateman}}, \bibinfo
  {author} {\bibfnamefont {K.}~\bibnamefont {Bongs}}, \bibinfo {author}
  {\bibfnamefont {S.}~\bibnamefont {Bose}}, \bibinfo {author} {\bibfnamefont
  {C.}~\bibnamefont {Braxmaier}}, \bibinfo {author} {\bibfnamefont
  {{\v{C}}.}~\bibnamefont {Brukner}}, \bibinfo {author} {\bibfnamefont
  {B.}~\bibnamefont {Christophe}},  \emph {et~al.},\ }\href@noop {} {\bibfield
  {journal} {\bibinfo  {journal} {EPJ Quantum Technology}\ }\textbf {\bibinfo
  {volume} {3}},\ \bibinfo {pages} {1} (\bibinfo {year} {2016})}\BibitemShut
  {NoStop}%
\bibitem [{\citenamefont {Bateman}\ \emph {et~al.}(2015)\citenamefont
  {Bateman}, \citenamefont {McHardy}, \citenamefont {Merle}, \citenamefont
  {Morris},\ and\ \citenamefont {Ulbricht}}]{bateman2015existence}%
  \BibitemOpen
  \bibfield  {author} {\bibinfo {author} {\bibfnamefont {J.}~\bibnamefont
  {Bateman}}, \bibinfo {author} {\bibfnamefont {I.}~\bibnamefont {McHardy}},
  \bibinfo {author} {\bibfnamefont {A.}~\bibnamefont {Merle}}, \bibinfo
  {author} {\bibfnamefont {T.~R.}\ \bibnamefont {Morris}}, \ and\ \bibinfo
  {author} {\bibfnamefont {H.}~\bibnamefont {Ulbricht}},\ }\href@noop {}
  {\bibfield  {journal} {\bibinfo  {journal} {Sci. Rep.}\ }\textbf {\bibinfo
  {volume} {5}},\ \bibinfo {pages} {8058} (\bibinfo {year} {2015})}\BibitemShut
  {NoStop}%
\bibitem [{\citenamefont {Riedel}(2013)}]{PhysRevDRiedel}%
  \BibitemOpen
  \bibfield  {author} {\bibinfo {author} {\bibfnamefont {C.~J.}\ \bibnamefont
  {Riedel}},\ }\href {\doibase 10.1103/PhysRevD.88.116005} {\bibfield
  {journal} {\bibinfo  {journal} {Phys. Rev. D}\ }\textbf {\bibinfo {volume}
  {88}},\ \bibinfo {pages} {116005} (\bibinfo {year} {2013})}\BibitemShut
  {NoStop}%
\bibitem [{\citenamefont {Geraci}\ \emph {et~al.}(2010)\citenamefont {Geraci},
  \citenamefont {Papp},\ and\ \citenamefont {Kitching}}]{PhysRevLettGeraci}%
  \BibitemOpen
  \bibfield  {author} {\bibinfo {author} {\bibfnamefont {A.~A.}\ \bibnamefont
  {Geraci}}, \bibinfo {author} {\bibfnamefont {S.~B.}\ \bibnamefont {Papp}}, \
  and\ \bibinfo {author} {\bibfnamefont {J.}~\bibnamefont {Kitching}},\ }\href
  {\doibase 10.1103/PhysRevLett.105.101101} {\bibfield  {journal} {\bibinfo
  {journal} {Phys. Rev. Lett.}\ }\textbf {\bibinfo {volume} {105}},\ \bibinfo
  {pages} {101101} (\bibinfo {year} {2010})}\BibitemShut {NoStop}%
\bibitem [{\citenamefont {Yin}\ \emph {et~al.}(2013)\citenamefont {Yin},
  \citenamefont {Li}, \citenamefont {Zhang},\ and\ \citenamefont
  {Duan}}]{PhysRevAYin}%
  \BibitemOpen
  \bibfield  {author} {\bibinfo {author} {\bibfnamefont {Z.}~\bibnamefont
  {Yin}}, \bibinfo {author} {\bibfnamefont {T.}~\bibnamefont {Li}}, \bibinfo
  {author} {\bibfnamefont {X.}~\bibnamefont {Zhang}}, \ and\ \bibinfo {author}
  {\bibfnamefont {L.~M.}\ \bibnamefont {Duan}},\ }\href {\doibase
  10.1103/PhysRevA.88.033614} {\bibfield  {journal} {\bibinfo  {journal} {Phys.
  Rev. A}\ }\textbf {\bibinfo {volume} {88}},\ \bibinfo {pages} {033614}
  (\bibinfo {year} {2013})}\BibitemShut {NoStop}%
\bibitem [{\citenamefont {Genoni}\ \emph {et~al.}(2015)\citenamefont {Genoni},
  \citenamefont {Zhang}, \citenamefont {Millen}, \citenamefont {Barker},\ and\
  \citenamefont {Serafini}}]{1367-2630-17-7-073019}%
  \BibitemOpen
  \bibfield  {author} {\bibinfo {author} {\bibfnamefont {M.~G.}\ \bibnamefont
  {Genoni}}, \bibinfo {author} {\bibfnamefont {J.}~\bibnamefont {Zhang}},
  \bibinfo {author} {\bibfnamefont {J.}~\bibnamefont {Millen}}, \bibinfo
  {author} {\bibfnamefont {P.~F.}\ \bibnamefont {Barker}}, \ and\ \bibinfo
  {author} {\bibfnamefont {A.}~\bibnamefont {Serafini}},\ }\href
  {http://stacks.iop.org/1367-2630/17/i=7/a=073019} {\bibfield  {journal}
  {\bibinfo  {journal} {New J. Phys.}\ }\textbf {\bibinfo {volume} {17}},\
  \bibinfo {pages} {073019} (\bibinfo {year} {2015})}\BibitemShut {NoStop}%
\bibitem [{\citenamefont {Wiseman}\ and\ \citenamefont
  {Milburn}(1993)}]{PhysRevLett.70.548}%
  \BibitemOpen
  \bibfield  {author} {\bibinfo {author} {\bibfnamefont {H.~M.}\ \bibnamefont
  {Wiseman}}\ and\ \bibinfo {author} {\bibfnamefont {G.~J.}\ \bibnamefont
  {Milburn}},\ }\href {\doibase 10.1103/PhysRevLett.70.548} {\bibfield
  {journal} {\bibinfo  {journal} {Phys. Rev. Lett.}\ }\textbf {\bibinfo
  {volume} {70}},\ \bibinfo {pages} {548} (\bibinfo {year} {1993})}\BibitemShut
  {NoStop}%
\bibitem [{\citenamefont {Wiseman}\ and\ \citenamefont
  {Milburn}(2010)}]{wiseman2010quantum}%
  \BibitemOpen
  \bibfield  {author} {\bibinfo {author} {\bibfnamefont {H.}~\bibnamefont
  {Wiseman}}\ and\ \bibinfo {author} {\bibfnamefont {G.}~\bibnamefont
  {Milburn}},\ }\href {https://books.google.co.uk/books?id=ZNjvHaH8qA4C} {\emph
  {\bibinfo {title} {Quantum Measurement and Control}}}\ (\bibinfo  {publisher}
  {Cambridge University Press},\ \bibinfo {year} {2010})\BibitemShut {NoStop}%
\bibitem [{\citenamefont {Beresnev}\ \emph {et~al.}(1990)\citenamefont
  {Beresnev}, \citenamefont {Chernyak},\ and\ \citenamefont
  {Fomyagin}}]{beresnev1990motion}%
  \BibitemOpen
  \bibfield  {author} {\bibinfo {author} {\bibfnamefont {S.}~\bibnamefont
  {Beresnev}}, \bibinfo {author} {\bibfnamefont {V.}~\bibnamefont {Chernyak}},
  \ and\ \bibinfo {author} {\bibfnamefont {G.}~\bibnamefont {Fomyagin}},\
  }\href@noop {} {\bibfield  {journal} {\bibinfo  {journal} {J. Fluid Mech.}\
  }\textbf {\bibinfo {volume} {219}},\ \bibinfo {pages} {405} (\bibinfo {year}
  {1990})}\BibitemShut {NoStop}%
\bibitem [{\citenamefont {Varga}\ and\ \citenamefont
  {T{\"o}r{\"o}k}(2000)}]{varga2000focusingnumerical}%
  \BibitemOpen
  \bibfield  {author} {\bibinfo {author} {\bibfnamefont {P.}~\bibnamefont
  {Varga}}\ and\ \bibinfo {author} {\bibfnamefont {P.}~\bibnamefont
  {T{\"o}r{\"o}k}},\ }\href@noop {} {\bibfield  {journal} {\bibinfo  {journal}
  {JOSA A}\ }\textbf {\bibinfo {volume} {17}},\ \bibinfo {pages} {2090}
  (\bibinfo {year} {2000})}\BibitemShut {NoStop}%
\bibitem [{\citenamefont {Cohadon}\ \emph {et~al.}(1999)\citenamefont
  {Cohadon}, \citenamefont {Heidmann},\ and\ \citenamefont
  {Pinard}}]{PhysRevLettCohadon}%
  \BibitemOpen
  \bibfield  {author} {\bibinfo {author} {\bibfnamefont {P.~F.}\ \bibnamefont
  {Cohadon}}, \bibinfo {author} {\bibfnamefont {A.}~\bibnamefont {Heidmann}}, \
  and\ \bibinfo {author} {\bibfnamefont {M.}~\bibnamefont {Pinard}},\ }\href
  {\doibase 10.1103/PhysRevLett.83.3174} {\bibfield  {journal} {\bibinfo
  {journal} {Phys. Rev. Lett.}\ }\textbf {\bibinfo {volume} {83}},\ \bibinfo
  {pages} {3174} (\bibinfo {year} {1999})}\BibitemShut {NoStop}%
\end{thebibliography}%

\clearpage
\newpage
\onecolumngrid

\begin{center}
\textbf{\large Supplementary Material for ``Parametric Feedback Cooling of Levitated Optomechanics in a Parabolic Mirror Trap''}
\end{center}

{\bf Abstract:} Here we give more details about the technical and theoretical details for the particle trap. We describe the interferometric method to detect the position of the particle and evaluate the resolution of this measurement. We describe the method to extract particle and trap parameters from the fitting to the theoretical model to the experimental power spectral density (PSD) and from scanning the trapping laser wavelength. We give details about the particle preparation, trapping and feedback cooling. We further show how to evaluate the numerical aperture (NA) of the parabolic mirror as used in the experiment. We discuss the analysis of noise effects in the trap and give more details about the implementation of the parametric feedback.


\section{S1 Extraction of parameters from fit to power spectral density (PSD)}
Here we describe how to extract nanoparticle parameters, such as mass $m$ and radius $r$, and parameters about the motion of the nanoparticle in the trap, such as damping of the particle motion $\Gamma_0$, from fitting to the measured power spectral density. Experimental data is directly recorded from the photodiode signal, which means the particle position is recorded in volts as function of time.

{\it Equation of Motion:} 
The time trace signal contains generally the motion of the particle in
\emph{x}, \emph{y} and \emph{z} degrees of motion of the particle, which we simplify to one dimension \emph{x}:
\begin{equation}
\label{eq:equationofmotion}
\ddot{x}(t)+\Gamma_{0} \dot{x}(t) +\frac{k_{0} +k_{fb}(t)}{m}x(t)= \frac{F_{\rm th}(t)}{m} 
\end{equation}

{\it Power spectral density (PSD) with gamma factor $\gamma$:}
We can write the PSD of the trapped nanosphere from
the above equation of motion. However, to be able to fit to the
experimental data it should be noted that the above is measured in
nm$^{\text{2}}$/Hz. Whereas experimental PSD of the recorded time trace is in
V$^{\text{2}}$/Hz, therefore the experimental PSD should look like as
\begin{equation}
\label{eq:fullpsd}
S_{x}(\omega) = \gamma^2\frac{k_B T_0}{\pi m}\frac{\Gamma_0}{([\omega_0 + \delta \omega]^2 - \omega^2)^2 + \omega^2[\Gamma_0 + \delta \Gamma_0]} ,
\end{equation}
where $\gamma$ is the conversion factor from volts to nanometers
in units of V/m. 

{\it Fitting gives A, B, C parameters in the PSD model:} 
We can simplify Eq.~(\ref{eq:fullpsd}) as
\begin{equation}
\label{eq:psdfit}
S_{x}^{exp} = \frac{A}{(B^2-\omega^2)^2+\omega^2 C^2} .
\end{equation}
We fit the experimental PSD with the simplified
Lorentzian according to Eq.~(\ref{eq:psdfit}), where $A:=\frac{\gamma^2 k_B T_0\Gamma_0}{\pi m}$,
$B:= \omega_0 + \delta \omega$ and $C:= \Gamma_0 + \delta \Gamma$ are free fit parameters. To obtain the particles mass $m$ we fit to the PSD of
a particle ($\delta\omega=0$, $\delta\Gamma=0$) at thermal equilibrium at $T_0=$300K. 

{\it Obtaining radius $r$:} 
Based on the fluctuation-dissipation theorem the damping of the
particle is $\Gamma_0$ and thus can be described as~\cite{beresnev1990motion}
\begin{equation}
\label{eq:fullGamma0}
\Gamma_0 = \frac{6 \pi \xi r}{m} \frac{0.619}{0.619+Kn}(1 +c_k),
\end{equation}
where $r$ is the radius of the trapped particle, $\xi=18.6\mu$Pa s is the viscosity of air, $c_k=(0.31Kn)/(0.785+1.152Kn+Kn^2)$ is a function of the Knudsen number
\begin{equation}
Kn = \overline{l}/p,
\end{equation}
where $\overline{l}$ is the mean free path of the gas particles at the given pressure $p$. We Taylor expand Eq.~(\ref{eq:fullGamma0}) for small Kn$^{\text{-1}}$ to first
order and rewrite for radius $r$,
\begin{equation}
\label{eq:radius}
r = 0.619 \frac{9 \pi \xi d_m^2}{\sqrt{2} \rho k_B T_0}\frac{p}{\Gamma_0} ,
\end{equation}
where $d_m$ is the diameter of air molecules and $\rho$ is the density of the nanoparticle. 

{\it Calculating the maximal possible position resolution of the detection method used:} Note that $p$ and $\Gamma_0$ are quantities measured and extracted from parameter $C$ of the Lorentzian fit,
respectively, thus are directly from experimental data. From \emph{r} it is
easily enough possible to obtain the mass of the trapped particle. 
From having obtained the mass of the particle we can go back to the
PSD fitting Eq.(\ref{eq:psdfit}), and rearrange to give the conversion
factor,
\begin{equation}
\label{eq:conversationfactor}
\gamma = \sqrt{\frac{A}{C}\frac{\pi m}{k_B T_0}}
\end{equation}

The noise equivalent power (NEP) of a detector characterises the resolution of a detector. For the balanced photodiode detectors used the $NEP_{det} = 70nV/\sqrt{Hz}$. Equivalently, we can work out the position resolution of our setup
\begin{equation}
\label{eq:positionsensitivity}
S_{x,min} = \frac{NEP_{det}}{\gamma} .
\end{equation}
A typical value for $S_{x,min} $ is 17fm$/\sqrt{Hz}$, which is the minimum achievable in our system. 

The current experimental position resolution, $S_{x,exp} = 0.53 pm/\sqrt{Hz}$, is limited by
the noise floor which currently is at $NEP_{exp} = 2\mu V/\sqrt{Hz}$ as analysed from experimental data as in Fig.1a). Parameters as extracted from the fit to the PSD shown in Fig.1 are summarised in Tab.\ref{tab:fittingtable}.
\begin{table}[htb]
\centering
\begin{tabular}{l|r|l|l}
Quantity & Typical & Units & Error\\
 & Value &  & ($\pm \%$)\\
\hline
Pressure, $p$ & $7 \times 10^{-2}$ & mbar & 15\\
Collisional Damping Rate, $\Gamma_0$ & 400 & Hz & 3\\
Radius, $r$ & 75 & nm & 18\\
Mass, $m$ & $3 \times 10^{-18}$ & kg & 21\\
Conversion Factor, $\gamma$ & $4 \times 10^{6}$ & V/m & 34\\
C.o.M Temperature, $T_0$ & $\sim$ 300 & K & \\
Q factor, & $2 \times 10^{3}$ &  & 4\\
S$_{\text{x,min}}$ & $17 \times 10^{-15}$ & m$/\sqrt{Hz}$ & 34\\
S$_{\text{x,exp}}$ & $53 \times 10^{-11}$ & m$/\sqrt{Hz}$ & 34\\
\hline
 &  &  & \\
\end{tabular}
\caption{\label{tab:fittingtable}Table shows the parameters obtained via fitting to a PSD of an uncooled 75nm radius nanosphere trapped at 7$\times$10$^{-2}$mbar. The values given here correspond to the frequency domain trace shown in Fig.1.}
\end{table}

\begin{figure}
    \centering
    \includegraphics[width=0.9\textwidth]{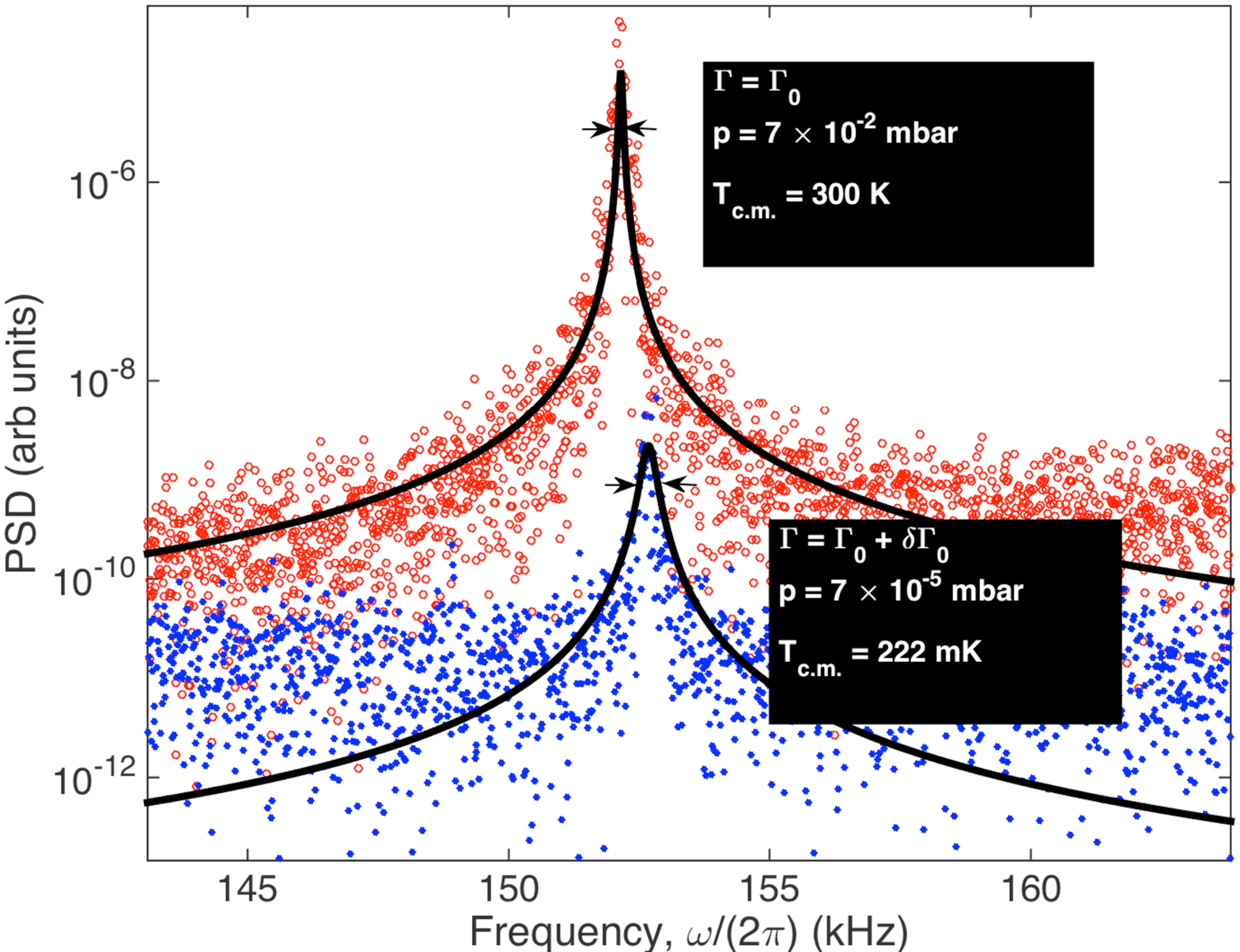}
    \caption{ \textbf{Temperature evaluation from fits to PSD.} The hollow points show the PSD for the z-motion with no feedback applied. The temperature is therefore assumed to be 300K. The solid points show the PSD for the z-motion with feedback turned on. The temperature of the z-motion is evaluated by fitting a Lorentzian to the experimental data points and has been evaluated to be 222mK. It can be seen that the peak approaches the noise floor, and a fit is still possible.The laser power used for this figure is higher than that in the main text and Fig.3, hence the Z frequency peak occurs at a higher frequency.}
    \label{fig:temperature}
\end{figure}

{\it Evaluation of mechanical quality factor $Q_m$:} The PSD is fitted by the Lorentzian according to Eq.(\ref{eq:fullpsd}) and the full width at half maximum $\Delta\omega$ is compared to the mean frequency $\omega_0$ of the motion in the respective direction: $Q_m=\omega_0/\Delta\omega$. $Q_m$ for the 300K peak in Fig.\ref{fig:temperature} is given in Tab.\ref{tab:fittingtable}. Assuming the motion of the particle is dominated by gas collisions, then the width of the PSD is determined by the collisional damping rate $\Gamma_0$ and we can write: $Q_m=\omega_0 / \Gamma_0$. We apply this analysis to the lower graph in Fig.\ref{fig:temperature} and find $Q_m=$5$\times$10$^{7}$.  

{\it Evaluation of the temperature of the motion from Lorentzian fits:}  We assume that the particle without feedback is at thermal equilibrium with the surrounding gas at $T_0 = 300$K, which is confirmed by our experimental study for various trapping laser powers. Fitting the Lorentzian according to Eq.(\ref{eq:fullpsd}) to a PSD of the particles motion for a particle at thermal equilibrium with the back ground gas we are able to extract $\gamma^2k_B T_0/ \pi m$. Using this we are able to extract the damping parameters $\Gamma_0$ and $\delta\Gamma$ from a fit to a PSD of particle motion for a cooled nanoparticle as shown in Fig.(\ref{fig:temperature}). Finally we can obtain $T_{cm}$ by using 

\begin{equation}
\label{eq:tempcal}
T_{cm}=T_{0} \frac{\Gamma_0}{\Gamma_0 + \delta\Gamma}.
\end{equation}

\section{S2 Evaluation of resolution of position detection from wavelength scan of trapping light:} Here we describe the detection of the particle position by analysing the optical interference pattern of the field which is Rayleigh scattered by the particle $E_{scat}$, which is re-collimated by the parabolic mirror and the diverging reference field $E_{div}$, which is diverging and therefore loosing intensity radiating away from the mirror.

{\it Determining the particle position resolution:} In our case the trapped nanoparticle modulates the trapping laser
field as the back-scattered laser light from the particle acquires a position-dependent phase shift. Although in reality stochastically driven, here for simplicity we model the movement of the particle in the trap as $z(t)=z_0 sin(\omega t + \alpha)$, with $\alpha$ is the phase and $z_0$ is the amplitude of the motion of the particle in the $z$-direction. 

{\it The scattered field:} The oscillating particle then Rayleigh scatters the trapping light in all directions, half of which is then collected and collimated in the opposite direction of the trapping laser by the
paraboloidal mirror. The modulation of the scattered light depending the movement of the particle can be written as
\begin{equation}
\label{eq:scatteringfield}
E_{scat} = E_{scat,0} e^{i \beta \sin(\omega_{0} t)},
\end{equation}
where we can write $\beta = z_0\partial_z\alpha$. We work out the phase of the motion of the
particle to be $\alpha = kz - arctan(z/z_R)$, with $k$ is the spring constant and $z_R$ the Rayleigh distance of the optics. Therefore $\partial\alpha
= k-1/z_R$, assuming $z/z_R$ is small, which results in
\begin{equation}
\label{eq:betafactor}
\beta = kz_0 - \frac{z_0}{z_R}.
\end{equation} 

{\it The diverging reference field:} We represent the reference field as the
diverging field produced by the focussing of the trapping laser by the
paraboloidal mirror, represented as
\begin{equation}
\label{eq:referencefield}
E_{div} = E_{div,0} e^{i\theta},
\end{equation}
where $\theta = 2fk + \pi$, where \emph{f} is the focal length of the
mirror, and $\pi$-term is introduced due to the Gouy phase shift. 

{\it Interference of both fields, homodyne detection:} The combination of both the scattering field and the reference field
forms part of our homodyne detection scheme, which can be written as
\begin{equation}
\label{eq:intensity}
\begin{split}
I(t) & \approx  |E_{scat} + E_{div}|^2 \\
     & \approx E_{div,0}^2 + E_{scat,0}^2 + \\
& 2 E_{div,0}E_{scat,0}[\sin(\theta)\sin(\beta \sin(\omega_{0} t)) + \cos(\theta)\cos(\beta \sin(\omega_{0} t))] .
\end{split}
\end{equation}
Using the Jacobi-Anger identity we can expand the Bessel functions to leading order
\begin{equation}
\label{eq:intensityexpanded}
\begin{split}
I(t) & \approx E_{div,0}^2 + E_{scat,0}^2 \\
& + 2 E_{div,0}E_{scat,0}[\sin{(\theta)} \times J_0(\beta) + \sin{(\theta)} \times 2J_1(\beta)\sin{(\omega_{0} t)} \\
& + \cos{(\theta)} \times 2J_2(\beta)\sin{(2\omega_{0} t)}],
\end{split}
\end{equation}
where J$_{\text{n}}$($\beta$) is a Bessel function of the first kind, with
\emph{n}=(0,1,2\ldots{}). For the leading order in $\beta$, we can write $J_0(\beta)\approx$1, and  $J_1(\beta)$ and
$J_2(\beta)$ refer to the amplitude of the first and second harmonic of the particle motion along a given axis. The ratio of these two
amplitudes gives us the a means by which we can extract $z_0$. By
expanding the Bessel functions to their first order we have
\begin{equation}
\label{eq:rfactor}
\begin{split}
\varrho & = \frac{J_2(\beta)}{J_1(\beta)} = \frac{\beta^2/8}{\beta/2},\\
  & = \frac{1}{4}\beta.
\end{split}
\end{equation}
The ratio of these amplitudes we obtain experimentally. We also know that the amplitudes of these terms depend on the wavenumber
\emph{k}. By changing the wavelength we can make the amplitude of one harmonic
more prominent than the other, while the sum of the square of the two amplitudes should stay  constant.  Since we also know z$_{\text{0}}$ in terms of volts through our recorded signal we can then produce an alternative means to obtain $\gamma$, the conversion factor from Eq.(\ref{eq:conversationfactor}). We can then obtain mass $m$ and radius $r$ of the particle without any knowledge of
the pressure $p$ and damping factor $\Gamma_0$. The method is also transcend any assumption about the theoretical kinetic model being used. 

\begin{figure}
    \centering
    \includegraphics[width=0.8\textwidth]{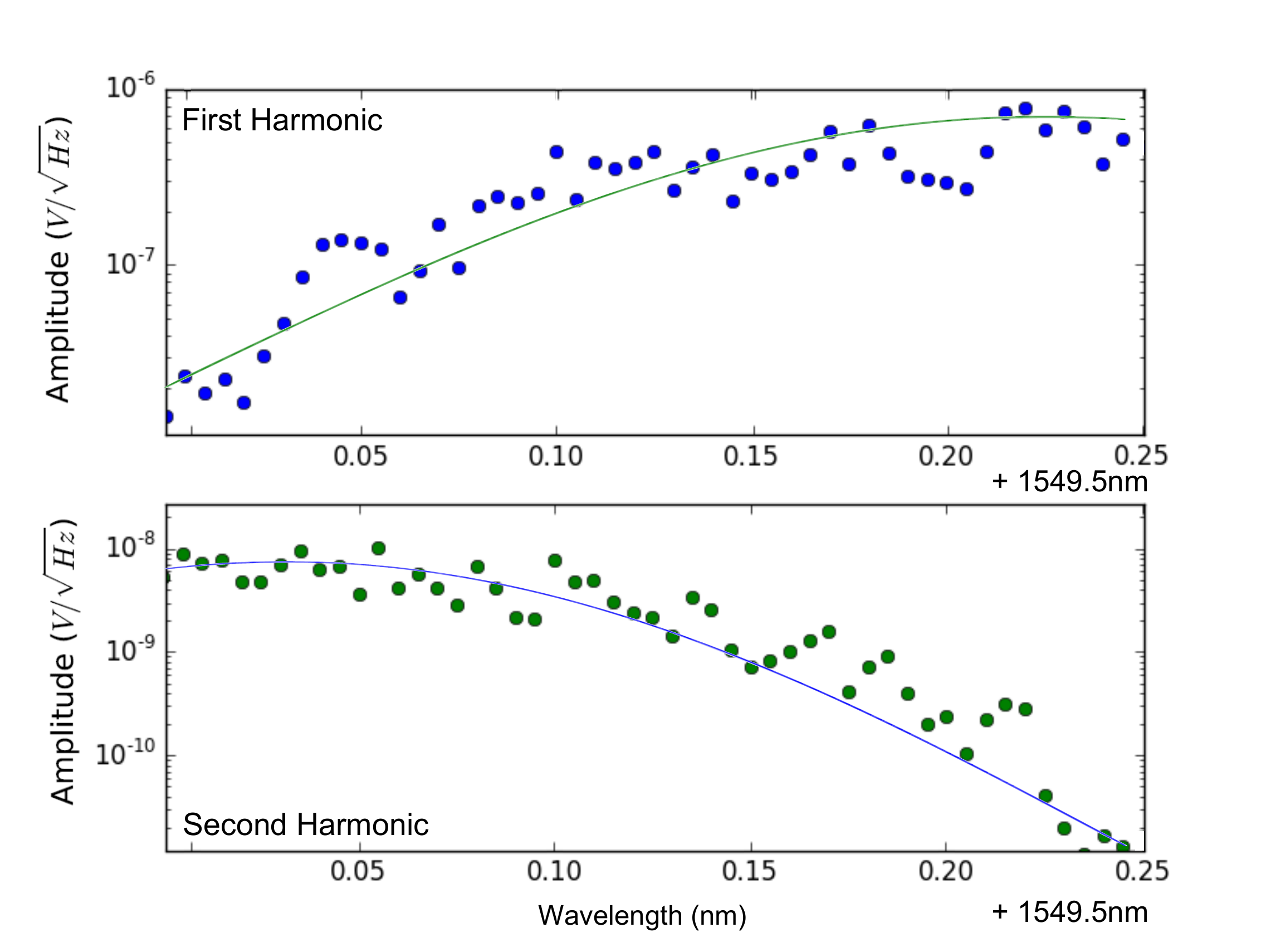}
    \caption{ {\bf Wavelength scan to measure relative amplitude change in the first and second harmonic of the trap frequency for the z-peak.} The relative amplitude change depending on the wavelength of the trapping laser is used to evaluate the resolution for the detection of the particle position. For a 60nm diameter silica particle, trapped at $1\times 10^{-2}$mbar we can observe how the amplitude of the first order peak (top panel, fitted with $\sin{(\theta)} 2J_1(\beta)\sin(\omega_{0} t)$) and second order peak (lower panel, fitted with $\cos{(\theta)} 2J_2(\beta)\sin{(2\omega_{0} t)}$), changes as the wavelength of the laser if varied to extract a parameter independent position resolution. }
   \label{fig:wavescan}
\end{figure}

{\it Experimental implementation of wavelength scan:} The parabolic mirror used has a focal distance of $f=3.1$ mm. By varying the trapping laser wavelength in the range 1545nm to 1555nm in steps of 5pm, with this we are able to vary the phase $\theta$ by $5.2\pi$ in steps of $0.03\pi$. 

{\it Results:} Fig.3c) by comparing the change in amplitude of the first and second order $z$ motion peaks we calculate the value for $\beta$ in Eq.(\ref{eq:betafactor}). From this we extract the maximum position of particle motion in the $z$ direction, $z_{0}=119\pm10$nm. From equipartition theorem: $m= k_{b}T_{cm}/\omega_{0}^{2}z_{0}^{2} $, with $T_{cm} = 300$K we can obtain a pressure independent measure of the particles mass m$=3\pm0.5\times10^{-19}$kg. Assuming a spherical particle the particle has a radius r$=30$nm$(\pm2)$nm. Finally using $\gamma = V/z_{0}$ where V is the voltage detected on our photodiode, and equation 12 it is possible to calculate the the position resolution of our set up to be $S_{x,exp}=200\pm20 fm/\sqrt{Hz}$.

This wavelength scan method to evaluate the position resolution is more accurate than the method used in supplement S1. Applying the methodology in S1 results in larger errors originating from large uncertainty in the measurement of the pressure in the vacuum chamber, by circumventing pressure readings and exclusivity using the measurements of the of the optical intensity measured by our detector.In Principle this method can be used to measure any spherical particle without knowing its density.


\section{S3 Particle preparation and trapping}

{\it Nanoparticle preparation and trap loading:} Nanoparticles loaded into the optical trap at atmospheric pressure via the use of an Omron micro-Air nebuliser. The nebuliser disperses droplets containing individual nanoparticles. The trapping region was evacuated by a turbo-molecular pump (Pfeiffer, 70l/s) to as low as 1$\times$10$^{-6}$mbar with the particle in the trap. The nebuliser is loaded with a suspension of aqueous liquid containing the nano-particles. The suspension used is prepared by diluting a suspension of 26nm -160nm (diameter) silicon dioxide nano-particles in deionized water using micro pipettes. The suspension is then sonicated for 15 minutes (37kHz, 300K in an ultrasound bath) before being loaded into the nebuliser. A concentration of 76,000 particles per ml was found to work well experimentally when the particles where dispersed into the vacuum chamber. Once a particle is trapped we begin to evacuate the chamber. We tried particles from many different suppliers and manufacturers, but had best trapping and pumping results with particles from the companies micro-particles GmbH and Corpuscular. The trapping success was increased significantly by turning the feedback stabilisation on with previously known feedback parameters.

{\it Experimental setup of trap:} The optical trap is generated with light from a stabilised fibre laser ($\lambda=1550$nm, NKT Koheras Basik CI5, 40mW). The output of the laser is intensity modulated by a AOM (PhotonLines, 80MHz) according to a feedback signal (see supplement S6 for details). The light then seeds an erbium doped fibre amplifier (EDFA, NuPhoton) to a maximum power of 1W. The light is focused by a high numerical aperture (NA=0.995) parabolic aluminium mirror which is mounted in a vacuum chamber. We trap silica (SiO$_2$) nanoparticles at vacuum as low as 1$\times$10$^{-6}$mbar and at trap frequency $\omega_0$. This trap frequency can be controlled by varying the laser power and hence the optical spring constant to achieve a $\omega_0$ in the range 10kHz to 300kHz. The particle is a driven dipole and hence radiates a field, a backwards propagating fraction of which is collected and collimated by the same parabolic mirror. Interference between this back scattered field $E_{scat}$, and the highly divergent field which passes without interacting with the particle through the focus  $E_{div}$ (the local oscillator reference field), provides interferometric position resolution (see Fig.~1d) in the main paper). By allowing the reference field to diverge, we make the reference field amplitude comparable with that scattered by the particle, giving a large modulation visibility at the detector. This results in a high particle position resolution.



\section{S4 The parabolic mirror and calculating its NA}
The mirror has been precision machined directly from aluminium by the company Symonds Mirror Technology with a surface tolerance of 15nm. There is no optical or protective coating attached to the aluminium surface. The parabolic mirror as a reflective high NA optic is a cheap alternative to lens optics. The mirror is easy to implement and to us at ultra-high vacuum, which is more difficult for high NA objective lenses. Another advantage, the mirror does not have chromatic effects, which make the position of the focal point independent of the wavelength used. The latter might be interesting for multi-wavelength spectroscopy and manipulation techniques. 

\begin{figure}
    \centering
    \includegraphics[width=0.3\textwidth]{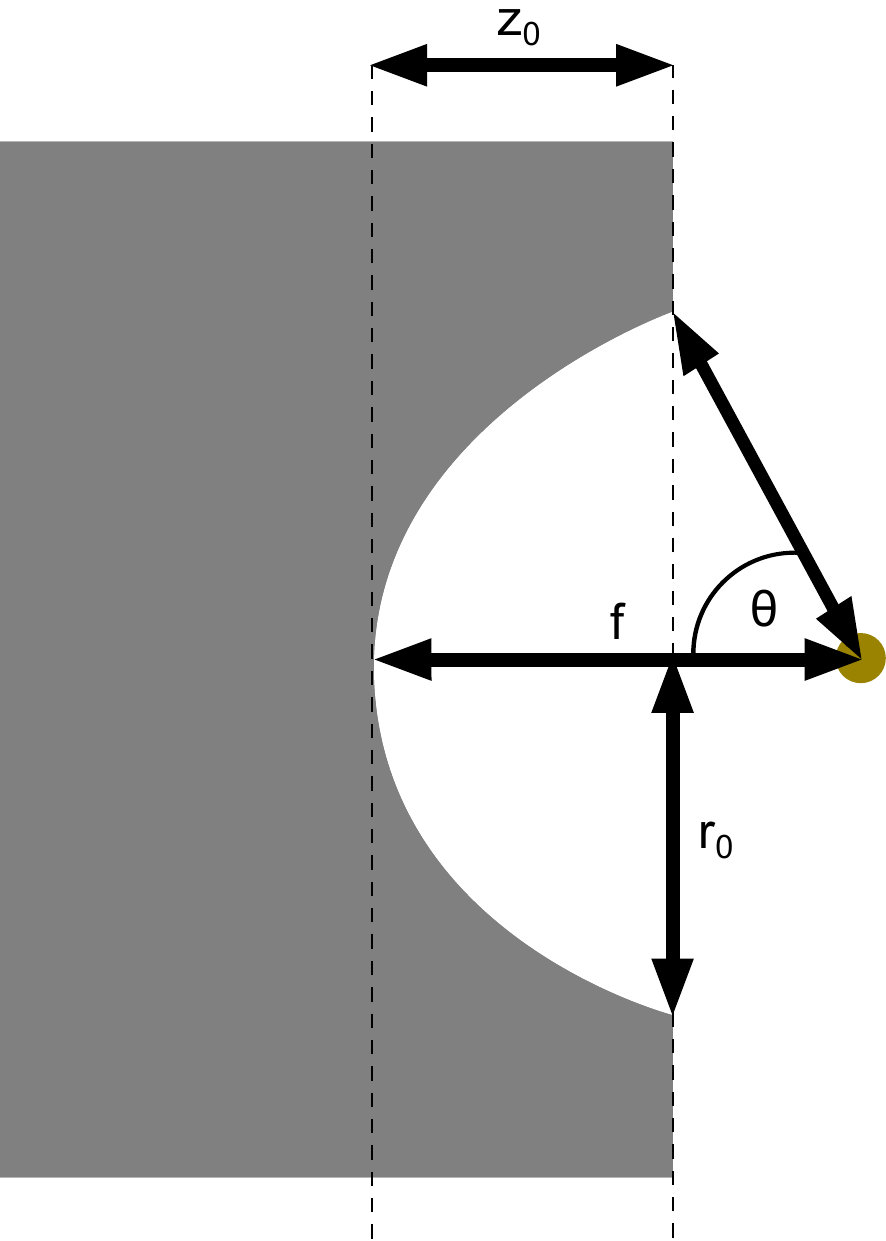}
    \caption{ \textbf{Geometry of the parabolic mirror.} The polarisable particle is trapped in the diffraction limited focal point of the parabolic mirror. We have used a number of different mirror designs though out the experiment, which differ in the working distance, the focal point with respect to the plane surface of the mirror and the numerical aperture NA. The more of the paraboloid used for the mirror, the higher the NA. The NA can also be varied/reduced for a given mirror by modification of the waist of the laser light incident on the mirror: If the light spot is smaller than the machined paraboloid, the the NA is less than the maximum. Typical mirrors, as used here, have a NA$\approx$1.}
    \label{fig:MirrorDiagram}
\end{figure}

{\it Evaluation of the numerical aperture:} The geometry of the mirror trap is illustrated in Fig.(\ref{fig:MirrorDiagram}). The numerical aperture  (NA) is defined as the light acceptance cone of the parabolic mirror. As the scattered light from a trapped particle diverges from the focus, we consider the solid angle at the distance $z_{0}$ as a fraction of the maximum angle of acceptance $2\pi$. Thus allowing us to write~\cite{varga2000focusingnumerical}:
\begin{equation} 
 \rm NA=\int_0^\theta\sin\theta'd\theta'=1-\cos\theta.
 \end{equation}
  
The general paraboloid function is $z=r^2/(4f)$, where $z$ is the height above the bottom of the mirror, $r$ is the radius of the paraboloid at $z$, and $f$ is the focal length. For the paraboloid as shown in Fig.\ref{fig:MirrorDiagram} at the maximum radius $r_0$ and at the corresponding height  $z_0$, with $z_0\leq f$, the angle $\theta$ between the optical axis and the edge of the paraboloid is given by

\begin{equation}
  \tan(\theta)=\frac{r_0}{(f-z_0)}=\frac{r_0}{f-r_0^2/(4f)}.
\end{equation}

We can thus define the NA of the parabolic mirror to be
\begin{equation}
\rm NA = 1-\cos \bigg( \arctan \bigg[ \frac{r_0}{f-r_0^2/(4f)} \bigg] \bigg).
\end{equation}

The mirror used for the majority of experiments in this study, had a focal length of $f$=3.1mm and $r_0$=12.7mm, which gives a NA of 0.995.


\section{S5 Evaluation of noises in the experiment by Allan deviation}
We analyse the noise in the experiment using Allan variance evaluation of the frequency dependent noise in the signal of the mechanical harmonic oscillator. The noise analysed contains information about fluctuation of amplitude, frequency and phase of the harmonically oscillating motion of the trapped particle.

Allan variance $\sigma_j^2$ can be depicted in the following steps.: 1) Take a long sequence of time domain data, $z(t)$. We choose z-direction of motion as this is the most sensitive in the experiment.
2) Chop signal in sequences according to time parameter $\tau$ and average the signal within $\tau$. 3) Repeat for various values of $\tau$ and plot according to:
\begin{equation}
\label{eq:allanvariance}
\sigma_j^2 = \frac{1}{2(n-1)} \sum_{i} \left(z_{i+1}(\tau_j) - z_{i}(\tau_j)\right)^2 .
\end{equation}
We will use Allan deviation which is given by $\sigma_j$. We repeat the procedure for variation of accessible experimental parameters: background pressure, feedback modulation depth $\eta$, and feedback phase. We find [in agreement with modelling] that the harmonic motion of the trapped particle is dominated by noise from collisions with background gas. Laser noise amplitudes are small compared to the collisional noise. 

{\bf Allan Deviation: Pressure Variation.}
Having created a means by which we can produce the Allan variance/deviation; we can start to compare at different parameter
variations. We are concerned specifically in our trapping frequency range, namely 10 kHz to 300 kHz, indicated as grey area in the Allan deviation figures.
\begin{figure}[htb]
\centering
\includegraphics[width=0.75\textwidth]{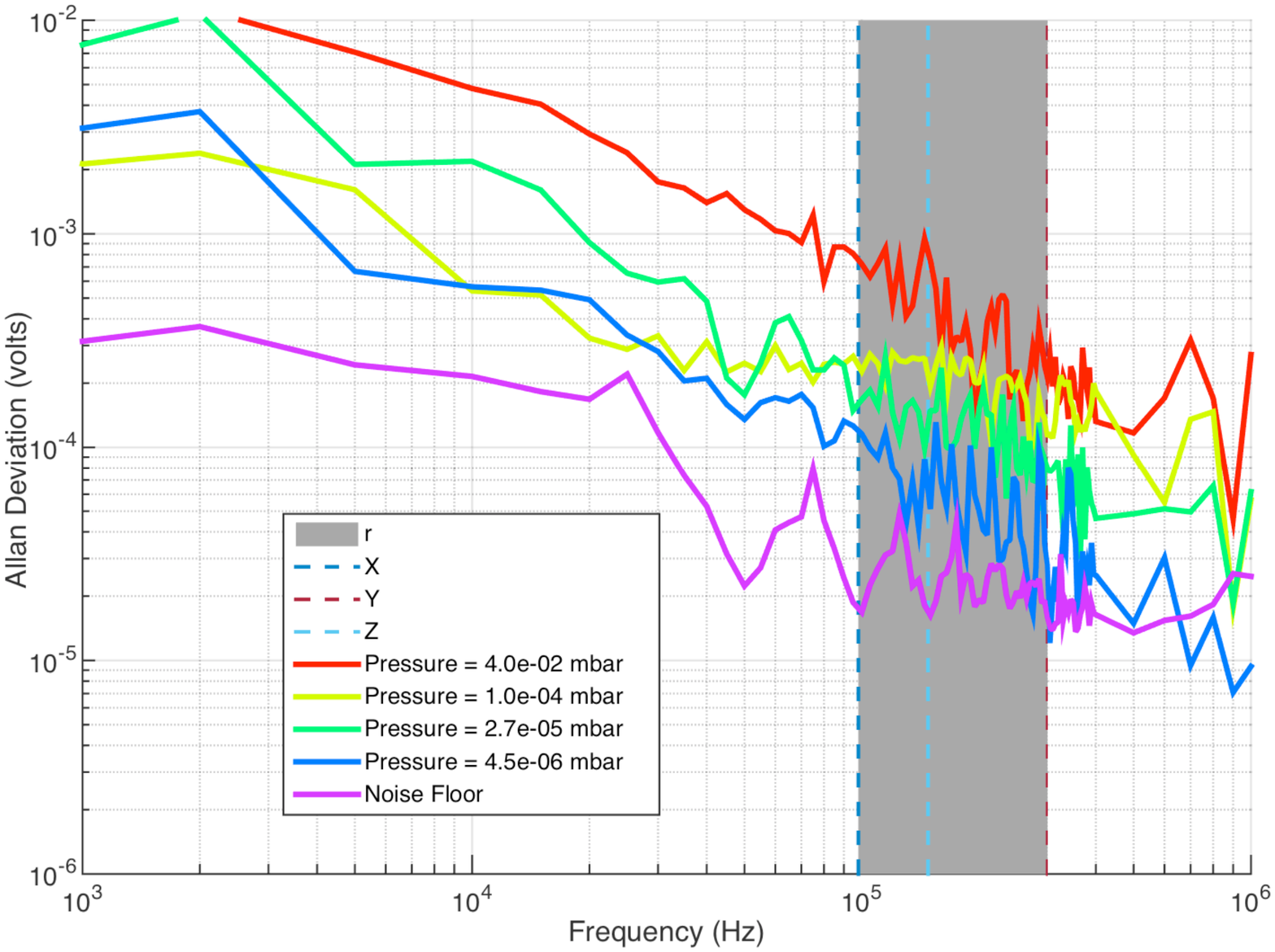}
\caption{\label{fig:avar_pressure}
{\bf Noise from pressure variation:}Allan variance of 150nm particle as it is cooled with constant feedback parameters but at different background pressure in the vacuum chamber. The different colour lines indicated Alan deviation at different pressures. The grey are indicated the frequency range of the mechanical oscillation of the trapped particle where the noise couples directly to the motion. Noise at the frequencies of $x,y,z$-motion, at along the vertical dashed lines, will need to be reduced as much as possible to realise a extreme high Q oscillator.}
\end{figure}
It is apparent from Fig.\ref{fig:avar_pressure} that by decreasing pressure the Allan deviation is decreasing and approaching the noise floor of the experiment. The noise floor was taken with no particle in the trap, but with the laser system, detectors and detection electronics in usual operation mode. It represents the minimum noise level possible at the given configuration of the experiment. The clear trend in pressure dependency toward the noise floor indicates that random gas collisions are the dominating noise source for the motion of the trapped particle. This is in agreement with the theoretical modelling described in the main text.

Each data string used for this Allan deviation noise analysis was taken for the duration of one second, which is a comparably short time. Therefore we do not expect to cancel [average out] any random noises in this measurements.
Data has not been taken at conditions and durations to perform a specifically low noise measurement. The Allan noise plots are meant to illustrate the usual noise level in a typical experiment, where there is clearly a lot of room for noise reduction procedures.

{\bf Allan Deviation: Feedback Modulation Depth Variation.}
Here we vary the feedback modulation depth $\eta$ while keeping all other parameters constant. 
\begin{figure}[htb]
\centering
\includegraphics[width=0.75\textwidth]{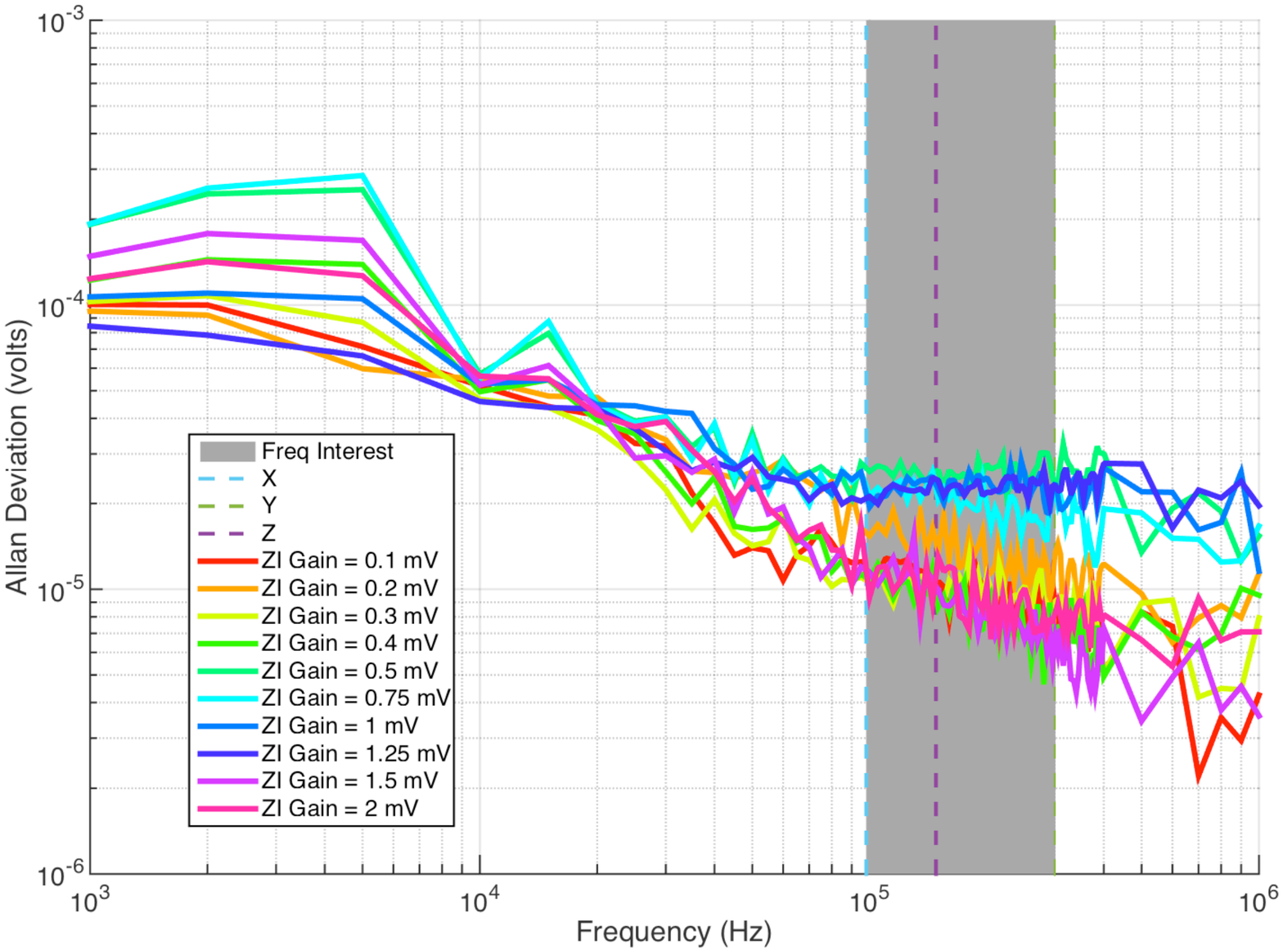}
\caption{\label{fig:avar_gain}{\bf Noise from feedback gain variation:} Allan deviation of the time-domain signal of the trapped particle motion in z-direction as it is cooled with increased modulation depth $\eta$ of the parametric feedback at constant pressure 5$\times10^{-6}$mbar and feedback phase $\phi$=70 degrees.}
\end{figure}
We can see that there is no clear $\eta$-dependency from the noise levels (see Fig.\ref{fig:avar_gain}) and the magnitude of the variation within the data set is much less compared to the observed variation for the pressure dependence. 

To further cool the particle a naive attempt could be to increase $\eta$ further. However we experimentally observe that after a certain gain factor threshold at a relative modulation depth of a few percent, we are unable to cool the motion of the trapped particle instead heating of the motion was observed. 

One could speculate that this is due to noise being introduced by the feedback while increasing the modulation depth. Instead this analysis, using Allan
deviation, indicates that this is not the case, as there is no obvious trend in the noise depending on $\eta$. Our analysis suggests that the reason for the non-trivial relation between $\eta$ and cooling temperature is within the parametric dynamics of the feedback itself.

\begin{figure}[htb]
\centering
\includegraphics[width=0.75\textwidth]{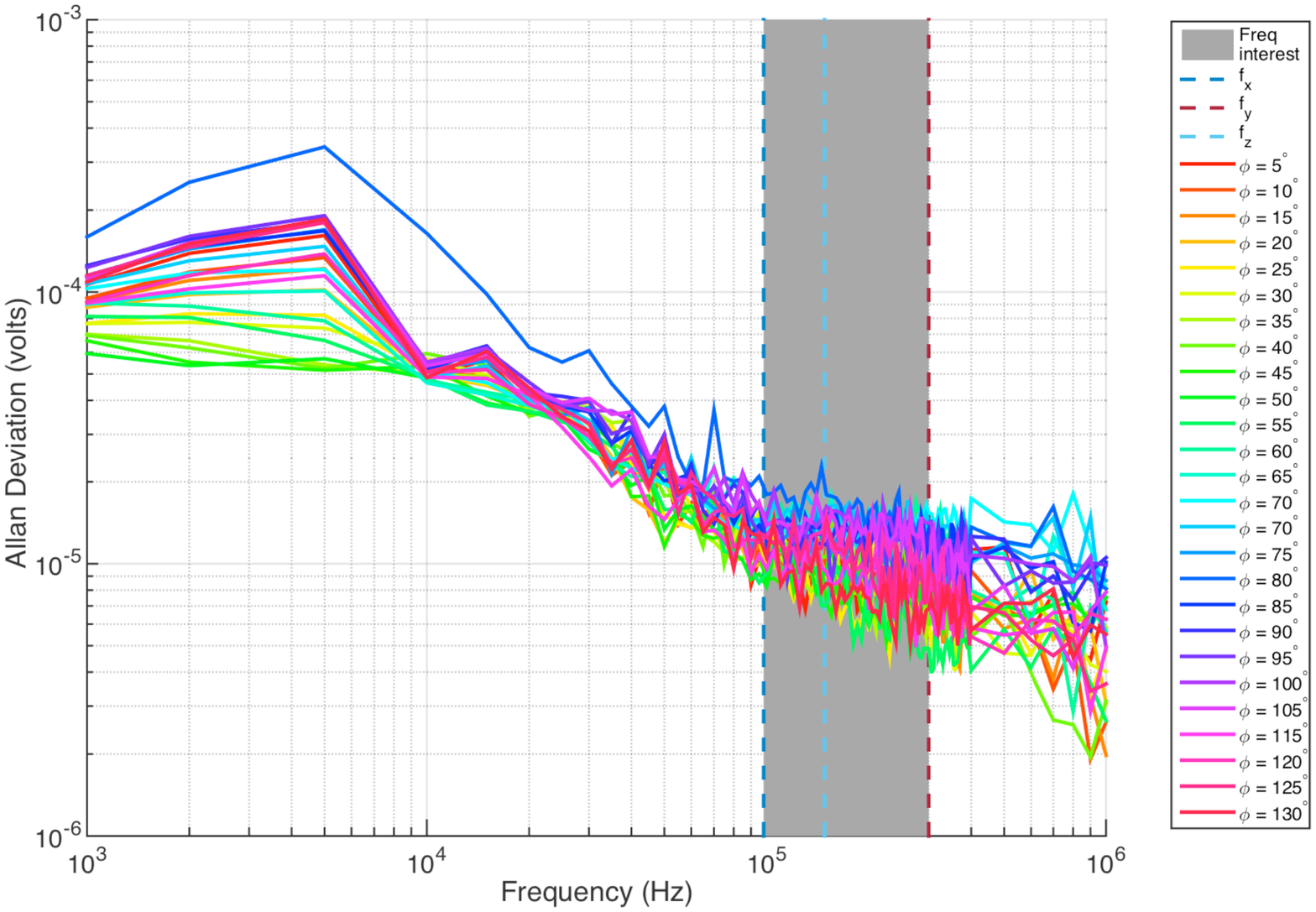}
\caption{\label{fig:avar_phase}{\bf Noise from feeddback phase variation:} Allan deviation for the backscattered light off a trapped 150nm particle. Pressure is constant at 5$\times 10^{-6}$mbar and the modulation depth is constant at $\eta$=0.39$\%$ . The phase of the feedback loop $\phi$ is varied as indicated in the legend of the plot.}
\end{figure}

{\bf Allan Deviation: Feedback Phase.}
As shown in Fig.3) (in the main text) modulation of the feedback phase $\phi$ affects the temperature of the particle's centre of mass motion in the trap. Once again the question we explore here is if this effect is related to noise introduced by the feedback. We plot Allan deviation as we change phase in Fig.\ref{fig:avar_phase} 
and find no clear relation between the feedback phase and the noise. In the main body of the paper we explain the observed heating and cooling effect as depending on the phase by an optical spring effect rather than related to noise.  

In summary, from Allan deviation of feedback modulation depth and of phase variation we conclude that no systematic noise effects are introduced to the motion of the particle by feedback. 


\section{S6 Feedback implementation}
\begin{figure}[h]
    \centering
    \includegraphics[width=0.7\textwidth]{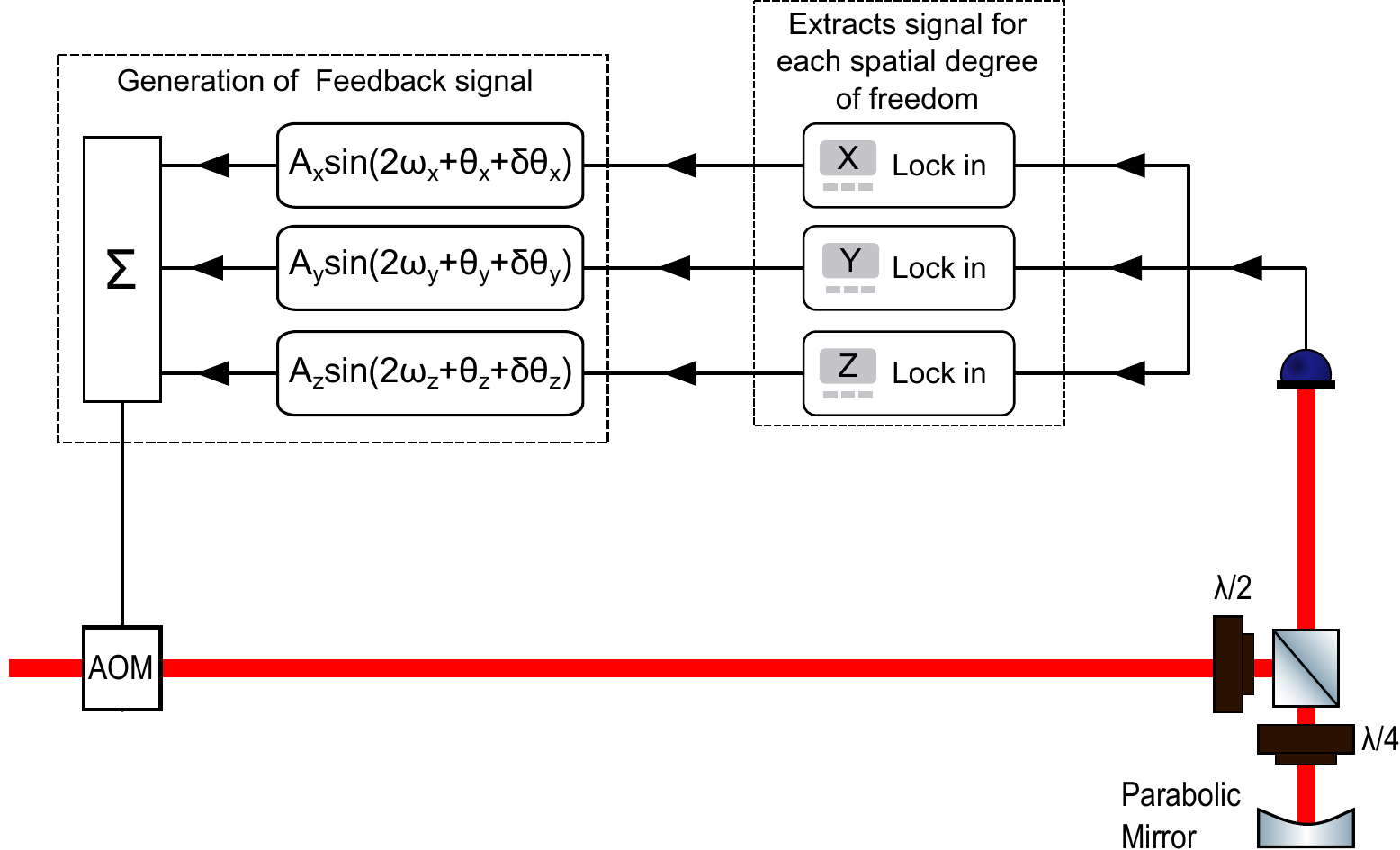}
    \caption{ \textbf{Feedback cooling scheme. }The scattered light from the particle is collected by a single photodiode. The detected signal can be considered a carrier wave containing three signals of differing frequency's corresponding to the trap frequency's of each spatial degree of freedom plus noise. This signal is then sent to three different lock in amplifiers (ZI HF2LI Lock-in Amplifier), where the signals are extracted for each spatial degree of freedom. In each of the lock in amplifiers the frequency $\omega $ and phase $\theta$ of the particle motion in one spatial degree of freedom is extracted. This information is then used to create a sine wave with amplitude $A$, twice the frequency of the trap frequency and a phase shift $\delta\theta $ relative to its measured phase, resulting in a feedback signal equivalent to $x\dot{x}$.    These frequency doubled and phase shifted signals for each spatial degree of freedom are added together before being sent to a AOM, to modulate the intensity of the laser beam.
}
    \label{fig:ExperimentalDiagram}
\end{figure}

While the first linear feedback scheme demonstrated in optomechanics for a cavity mirror setup~\cite{PhysRevLettCohadon}. the feedback was a phase shifted proportional feedback that applies a direct force via radiation pressure. The method here used is one of a parametric feedback and based on the modulation of the optical spring constant of the trap. Parametric feedback cooling takes advantage of parametric excitation, that is to say when a parametric oscillator is driven at twice the natural frequency the particles oscillation locks to the modulation phase resulting in amplification. By adjusting the phase of the driving at twice the frequency damping of the particles motion occurs. This means that to damp the motion of an optically trapped particle we need to modulate our laser intensity at twice the trap frequency and with a phase shift relative to that of the particle motion. The method used to achieve this experimentally can be seen in Fig.\ref{fig:ExperimentalDiagram}.  
To quantify the strength of the applied feedback we define the modulation depth $\eta$ of the optical feedback signal as $\eta = I_{fb}/I_{0}$where $I_{0}$ is the laser intensity without feedback and $I_{fb}$ is the amplitude of the feedback modulation. By adjusting the size of the modulation depth we can control the amount of damping we induce via feedback cooling and thus raise or lower the temperature of our trapped nanoparticle as desired (See Fig.2d in the main text).  This controlled heating and cooling of an optically trapped particle is a useful tool for the manipulation and the preparation of mechanical states of trapped nanoparticles. 


\section{S7 Derivation for connection of feedback modulation depth to underlying dynamics of the trapped particle}
Eq.(1) in the main text can be written in the following form
\begin{equation}
  \label{eq:2}
    \ddot{x}(t) + \Gamma_0 \dot{x}(t) + \omega_0^2
      x(t) = \frac{1}{m} (F_{th}(t) + F_{fb}(t)),
\end{equation}
where $F_{fb}(t) = - k_{fb}(t)x(t)$, considering the particles motion as $x(t)=A\sin(\omega_{0} t)$, we can write $F_{fb}$ in terms of the particles motion
\begin{equation}
  \label{eq:FP}
F_{fb}(t) = - \frac{2 \Delta k }{A^2 \omega_0} x(t)^2 \dot{x}(t).
\end{equation}
The modulation depth is defined as $\eta = \Delta k/k_{0}=\Delta k/m \omega_{0}^{2}$ allowing us to write the equation of motion in terms of our modulation depth
\begin{equation}
  \label{eq:2}
    \ddot{x}(t) + \Gamma_0 \dot{x}(t) + \omega_0^2
      x(t) = \frac{F_{th}(t)}{m} - \frac{2 \eta \omega_0}{A^2 } x^2(t) \dot{x}(t). 
\end{equation}
Making the slow-varying approximation $(x(t) = \bar{x} + \delta x(t) )$, which means that the mean of the position $\bar{x}$ is varying much slower with time as the position fluctuations $\delta x(t)$, we can write,
\begin{equation}
  \label{eq:6}
  \begin{split}
  \dot{x}(t)^2x(t) & = (\bar{x}(t) + \delta x(t))^2(\dot{\bar{x}}(t) + \delta \dot{x}(t))\\
  	& = (\bar{x}^2(t) + \delta x^2(t) +2\delta x(t)\bar{x}(t))(\dot{\bar{x}}(t) + \delta \dot{x}(t))\\
  	& =  \bar{x}^2(t)\dot{\bar{x}}(t) + \delta x^2(t)\dot{\bar{x}}(t) + 2\bar{x}(t) \dot{\bar{x}}(t) \delta x(t) \\
   & + \bar{x}^2(t) \delta \dot{x}(t) +  \delta x^2(t) \delta \dot{x}(t) + 2 \bar{x}(t) \delta x(t) \delta\dot{{x}}(t)
   \end{split}
\end{equation}
and retaining only the linear terms in $\delta x $, we write our feedback force Eq.(\ref{eq:FP}) as  
\begin{equation}
  \label{eq:6}
F_{fb}(t) = - \frac{2 \Delta k }{A^2 \omega_0} ( \overline{x}^{2}x(t) + \overline{x}^{2} \dot{x}(t)) ,
\end{equation}
noting $\overline{x} = A/2$, we can write the equation of motion in the following form

\begin{equation}
  \label{eq:3}
  \ddot{x}(t) + \Gamma_0 \left\{1 +\frac{\eta \omega_0}{2 } \right\}\dot{x}(t) + [\omega_0+\delta \omega]^2 x(t) = \frac{1}{m} F_{\rm th}(t).
\end{equation}
Taking the Fourier transform of the function and rearranging yields the following Power Spectral Density, $S_{x}(\omega)$,
\begin{equation}
  \label{eq:4}
  S_{x}(\omega) = \frac{2 k_b T_{0}}{\pi m}\frac{\Gamma_0}{([\omega_0+\delta \omega]^2 +
    \omega^2)^2 + \omega^2(\Gamma_0 + \eta \omega_0/2 )^2}
\end{equation}
\\
Knowing that the integral of a power spectral density can be written
as 
\begin{equation}
  \label{eq:5}
  \begin{split}
    \langle x^2 \rangle & = \int_{0}^{\infty} S_{x}(\omega) d\omega \\
               & = \frac{\Gamma_0 k_b T_{0}}{2 \pi m}
               \int_{-\infty}^{\infty}
               \frac{1}{([\omega_0+\delta \omega]^2+\omega^2)^2+\omega^2(\Gamma_0 +
                 \eta \omega_0/2 )^2} d\omega \\
               & = \frac{\Gamma_0 k_b T_{cm}}{m}
                \frac{1}{\omega_0^2 (\Gamma_0 + \eta \omega_0/2 ) }
  \end{split}   
\end{equation}
Using the equipartition theroem we note,
\begin{equation}
  \label{eq:6}
  \begin{split}
    \frac{1}{2}k_bT_{cm} & = \frac{1}{2} m \omega_{0}^2 \langle x^2
    \rangle \\
                                   & =  \frac{1}{2} m \omega_{0}^2
                                   \frac{\Gamma_0 k_b T_{0}}{ m}
                                   \frac{1}{\omega_0^2 (\Gamma_0 +
                                     \eta \omega_0/2 )}, \\
                                   \implies T_{cm} & = T_0 \frac{\Gamma_0}{\Gamma_0 + \eta \omega_0/2 }.
  \end{split}
\end{equation}
This shows an inverse relation to the modulation depth $\eta$ of the
feedback introduced to the system, in agreement with Eq.(9). From this we can see that if
$\eta \to \infty$ then $T_{cm} \to 0$. This of course is something we
do not observe experimentally. We find that as $\eta$ increases there
is plateau of the particles temperature and sometimes an increase in the temperature such that the particle can be lost from the trap. This we infer to be due to the particle experiencing a greater impulse at higher modulation depths and therefore, resulting in nonlinear response.


\section{S8 System Limitations}
In this section we discuss what causes limitations to cool towards the ground state in our system. The ability to cool a particle depends upon the resolution of which its position can be resolved. There are several sources of noise in the system which need to be minimized to produce the  highest possible position resolution including measurement uncertainty due to the random arrival of photons at the particle's location and measurement backaction due to momentum transfer from photons to the particle.

\subsection{S8.1 Detection resolution limits}
To examine whether the particle motion is resolvable in the ground state we consider the motion of the particle in the ground state given by 
\begin{equation}
  \label{eq:4}
\langle x_{ground} \rangle = \sqrt{\frac{\hbar}{m\omega_{0}}}.
\end{equation}
The size of the ground state for a 66nm silica particle with $\omega_{0}=2 \pi \times 100$kHz is $\langle x_{ground} \rangle = 6.5$pm. The detection resolution of our system has been evaluated in supplement S2 to be $S_{x,exp}=200\pm20 fm/\sqrt{Hz}$. For an integration time of 1 second we arrive at a resolution of $200\pm20 fm$. This resolution is sufficient to resolve 32 points per oscillation satisfying the Nyquist theorem. To cool a mechanical mode to near its ground state it is needed to resolve a zero-point motion $\Delta x$ at least within the (thermal) decoherence time. The mean thermal occupancy for an optically levitated nanoparticle is:
\begin{equation}
  \label{eq:4}
\langle n \rangle = \frac{k_{b} T_{n}}{\hbar \omega_{0}},
\end{equation}
where $T_{n}=k_{0}\langle x \rangle^{2}/k_{b}$, allowing us to write an expression for the  zero-point motion,
\begin{equation}
  \label{eq:4}
\Delta x  = \sqrt{\frac{k_{b}}{k_{0}}}(\sqrt{T_{n+1}}-\sqrt{T_{n}}).
\end{equation}
The zero-point motion $\Delta x = 2.6$pm and at $10^{-9}$mbar the thermal decoherence time is on the order of $10^{6} s$. Therefore for an integration time of one second a zero-point motion can be resolved and at the predicted ultra-high vacuum (and therefore low $\Gamma_0$) condition the detection resolution of our system should even allow for the spatial resolution of the ground state.

\subsection{S8.2 Detection efficiency limits}
Maximum detection efficiency occurs when $100\% $ of information carrying photons are collected by the detector in the system. To examine the detection efficiency in our system we must consider the fraction of light the particle sees of the laser light passing through the focus. The size of the laser focus is given by w$_{waist}= 2 \pi $w$_{0}$, where w$_{0}$ is the beam waist. While the particle will explore an area of w$_{particle} = \pi x_{0}y_{0}$. The fraction of light at the laser focus the particle sees is $\eta_{Area} = $w$_{particle}/$w$_{waist}$. The amount of light scattered by the particle is given by the Rayleigh scattering cross-section is given by, 
\begin{equation}
  \label{eq:4}
\sigma_{s} = \frac{2 \pi^{5}}{3}\frac{(2r)^{6}}{\lambda^4}\bigg(\frac{n^{2}-1}{n^2 +2}\bigg)^{2}.
\end{equation}
For a given laser power $P_{0}$ and radius of particle $r$, the power scattered by a nanoparticle within the trap is given by,
\begin{equation}
  \label{eq:4}
P_{scat} = \frac{\sigma_{s}}{2 \pi r^{2}} P_{0} \eta_{Area}.
\end{equation}
Therefore we can find the number of photons scattered in a given in a second is,
\begin{equation}
  \label{eq:4}
N_{scat} = P_{scatt}/h \omega_{0}.
\end{equation}
The percentage of photons collected  $\eta_{trans}$ can be calculated by accounting for the number of photons lost by absorption in the optical components and the amount of photons collected by the parabolic mirror, which collects $50\%$ of the scattered photons (a significant improvement on lens based traps). Combining this with the quantum efficiency of the detector $\eta_{Q}$ we can write an expression for the power detected as 
\begin{equation}
  \label{eq:4}
P_{det}=\eta_{Q}\eta_{trans}P_{scat}.
\end{equation}
The signal amplitude measured on the detector in volts is given by $I_{det}=P_{det} \times 1.0A/W \times G$. Where the transimpedance gain is given by $G \sim 10^{5}$V/A. Thus for the particle in the ground state with $\langle x_{ground}\rangle= 6.5$pm we expect about 30pV of electrical signal from the detector according to the estimated fraction of the Rayleigh scattered photons incident on the detector. The sensitivity of the balanced photo detector is $70nV/\sqrt{Hz}$ and the Zurich Instruments UHFLI lock-in amplifier has a sensitivity of 1nV. For the signal to therefore be detectable in the ground state an amplification/gain of about 2500 and an integration time of one second will be needed. We conclude that the detection efficiency is not putting a too strong constraint on the approach of the ground state.

\subsection{S8.3 Measurement noise in the system}
As the modulation depth of the feedback signal is increased an increase in the cooling rate is increased. However as the modulation depth exceeds $1\%$ the cooling rate starts to reduce and the particle temperature increases (as shown in Fig.(\ref{fig:moddepth})). This is suggestive of classical measurement noise, such as detection noise, electronic noise imparted by the FPGA system, amplifiers and the lock-in amplifiers in the system. Any such noises, if picked up by the measurement system, will be amplified by the feedback loop, which is the suspected reason for the observed effect of $\eta$ in Fig.(\ref{fig:moddepth}) and thus places a serious limit on the cooling rate that can be achieved in the system. However it is expected that the cooling rate achieved for modulation depths of $\eta\leq 1\%$ will be sufficient to cool to smaller $\bar n$ (ultimately only limited by the photon recoil rate, see below S8.4), if the background pressure and therefore also $\Gamma_0$ are reduced.

\subsection{S8.4 Photon Recoil Limit}
The laser light used to trap scattering off the particle imparts a momentum kick to the trapped particle, acting as an additional heating source. In our system the collisional damping from background gases is much larger than this effect. However as the particle is cooled and the temperature reduced a point will be reached where photon recoil will become an observable effect~\cite{Jain2016}. The photon recoil rate is given by,
\begin{equation}
  \label{eq:4}
\Gamma_{recoil}= \frac{1}{5}\frac{P_{scatt}}{m c^{2}}\frac{2 \pi c}{\lambda \omega_{0}}.
\end{equation}
For our system $\Gamma_{recoil}= 31.8$kHz, thus considering the phonon number at the steady state $n\delta\Gamma = 7000 \times 2\pi \times 500$Hz= 21MHz, we predict that at 24 phonons $\Gamma_{recoil}$ photon recoil will become a limiting factor in our system and will therefore prevent ground state cooling.

\end{document}